\documentclass[11pt]{article}
\usepackage{preamble}

\begin{document} 

\begingroup
\allowdisplaybreaks

\begin{titlepage}

\begin{flushright}
UWThPh 2026-8
\end{flushright}

\vskip 2cm
\begin{center}

{\Large{\textbf{Generalised Symmetries, Anomalies, and Maximal Branches \\[5pt] of 3d Chern–Simons Matter Theories}}}\\

\vspace{15mm}
{{\large 
Fabio Marino${}^a$, Francesca Pretto${}^{a,b}$, and 
Marcus Sperling${}^a$}} 
\\[5mm]
\noindent {${}^a$\em University of Vienna, Faculty of Physics, Mathematical Physics Group,\\
Boltzmanngasse 5, 1090 Vienna, Austria,}
\\[5mm]
\noindent {${}^b$\em University of Vienna, Vienna Doctoral School in Physics, \\ Boltzmanngasse 5, 1090 Vienna, Austria.}
\\[5mm]
Email: {\tt{fabio.marino@univie.ac.at}},\\ 
{\tt{francesca.pretto@univie.ac.at}},\\ 
{\tt{marcus.sperling@univie.ac.at}}
\\[15mm]
\end{center}

\begin{abstract}
The generalised symmetries, 't~Hooft anomalies, and their interplay with maximal branches are analysed for linear unitary three-dimensional $\Ncal\geq 3$ Chern--Simons Matter quiver theories realised in Type IIB brane configurations.
The 1-form symmetry, its self-anomaly and its maximal anomaly-free subgroup are determined independently from Wilson-line screening and from the string lattice. Fractional monopole endpoints of the generating Gukov--Witten defect are used to extract the anomaly data, while centre charges of integer monopoles determine the faithful 0-form symmetry group, its connected cover and possible 2-group structures. Mixed anomalies with non-abelian 0-form factors are found to cancel in the linear unitary theories considered. The effect of 1-form gauging on maximal branches is related to frozen D3-brane sectors, which flow to Chern--Simons TQFTs whose anyonic lines obstruct the required defect endpoint. This yields a criterion for affected branches and implies that at most two can be affected simultaneously. In such a case, a magnetic-quiver extension by non-simply laced edges is proposed, wherein the 1-form symmetry and its mixed anomaly with the branch isometry are reproduced.
\end{abstract}

\end{titlepage}

\clearpage
{\hypersetup{hidelinks} {\small{\tableofcontents}}}

\section{Introduction}
\label{sec:intro}
Generalised global symmetries extend ordinary 0-form symmetries, which act on local operators, to higher-form symmetries acting on extended operators \cite{Gaiotto:2014kfa,Cordova:2022ruw,Brennan:2023mmt}. Symmetries of different form degree may combine into higher-group structures, while their 't~Hooft anomalies govern whether they can be gauged. Determining these structures requires not only the symmetry algebras, but also their global forms, the spectrum of genuine and non-genuine operators, and the charges of the corresponding defects.

Three-dimensional Chern--Simons Matter (CSM) theories with $\Ncal\geq3$ supersymmetry \cite{Kao:1995gf,Schwarz:2004yj,Gaiotto:2007qi,Aharony:2008ug,Aharony:2008gk,Bagger:2007jr,
Gustavsson:2007vu} provide a natural setting in which these questions can be addressed. Chern--Simons couplings obstruct the screening of Wilson lines and thereby leave a discrete 1-form symmetry. At the same time, supersymmetry organises the moduli space of vacua into maximal branches. The resulting interplay between line operators, monopole operators and branch geometry is particularly accessible in Type IIB brane constructions, where the theories arise on D3 branes suspended between $(p,q)$ 5-branes \cite{Kitao:1998mf,Gauntlett:1997pk,Gauntlett:1997cv,Bergman:1999na}. Configurations containing only NS5 branes and one type of $(1,\kappa)$ 5-brane have enhanced $\Ncal=4$ supersymmetry \cite{Imamura:2008dt, Imamura:2008nn,Gaiotto:2008sd,Hosomichi:2008jd,Hosomichi:2008jb}; more general configurations yield $\Ncal=3$ theories. Their moduli spaces contain maximal branches associated with the different 5-brane types \cite{Assel:2017eun}, whose geometries are
encoded by magnetic quivers \cite{Marino:2025uub,Marino:2025ihk}.

The class considered here consists of linear unitary CSM quivers without fundamental hypermultiplets, equivalently without D5 branes. In these quivers, fundamental matter would screen the common fundamental Wilson-line class and trivialise the corresponding 1-form symmetry. The connected continuous 0-form symmetries relevant to the maximal branches arise from topological symmetries and may enhance to non-abelian groups in the infrared due to monopole operators \cite{Borokhov:2002cg,Gaiotto:2008ak,Bashkirov:2010hj}. Compared with the non-CS 3d $\Ncal=4$ quiver theories analysed in \cite{Bhardwaj:2022dyt,Bhardwaj:2023zix,Nawata:2023rdx}, and with the symmetry analyses of ABJ(M)-type and other 3d $\Ncal\geq3$ CSM theories \cite{Bergman:2020ifi,Beratto:2021xmn,Mekareeya:2022spm,Comi:2023lfm}, the additional question addressed here is how this higher-symmetry data is reflected in maximal branches and their magnetic quivers.

\paragraph{Results and strategy.}
The CSM theory is studied in two equivalent brane phases related by Giveon--Kutasov duality \cite{Giveon:2008zn}, or equivalently by the Hanany--Witten brane creation process \cite{Hanany:1996ie}. The \emph{factorised phase} makes the connected 0-form symmetry factors and the relevant monopole charges manifest, while the \emph{maximal branch phase} isolates the moduli and the residual topological sectors of a chosen maximal branch. The 1-form symmetry and its self-anomaly are independent of this choice of phase.

For a Lagrangian linear quiver with CS-levels $k_i$, screening of the fundamental Wilson lines gives $ \Gamma^{(1)}=\Z_g^{(1)}$, with $ g=\gcd( k_1,k_2,\ldots)$. The same result follows from the quotient of the $(p,q)$-string lattice by the strings that can end on the available 5-branes. The generator of $\Gamma^{(1)}$ is realised as a fractional central Gukov--Witten defect \cite{Gukov:2006jk,Gukov:2008sn}. Its fractional-monopole endpoint determines the self-anomaly, the maximal anomaly-free subgroup and the mixed anomaly with the topological 0-form symmetries. Integer monopole operators are instead used to determine the faithful connected group $\mathcal F$, its connected cover $F$, and the resulting 2-group structure. These constructions are developed uniformly for $\Ncal=4$ theories and extended to Lagrangian $\Ncal=3$ CSM quivers.

A direct relation between 1-form symmetry gauging and maximal branches is obtained from the D3 segments that remain frozen in a maximal branch phase. Their world-volume theories flow to Chern--Simons TQFTs. At a generic point of the branch, the 1-form generator becomes a diagonal combination of a sigma-model vortex line and anyonic Wilson lines of these ``frozen'' TQFTs. Since the latter admit no endpoint, any non-trivial frozen TQFT prevents the gauging from generating additional operators on that maximal branch. This gives a criterion for determining which branches are affected and implies, within the class analysed here, that at most two maximal branches can be affected simultaneously.

For an affected branch, an \emph{extension of the magnetic quiver} by non-simply laced edges is proposed. In the examples analysed, the extended quiver makes the 1-form symmetry manifest and reproduces its mixed anomaly with the maximal branch isometry. The brane analysis is also extended, where feasible, to non-Lagrangian $\Ncal\geq3$ theories engineered by generic $(p_i,q_i)$ 5-branes. In that setting the 1-form symmetry and the maximal branch isometries remain accessible, whereas the 1-form self-anomaly, the connected cover and the 2-group structure cannot presently be determined from explicit endpoint operators.

\paragraph{Structure of the paper.}
Section~\ref{sec:setup} fixes the field-theory and brane conventions. Section~\ref{sec:symm_anom} determines the 1-form, 0-form and 2-group structures and their anomalies for $\Ncal\geq3$ CSM theories. Section~\ref{sec:maxbranch_MQ} studies the effect of 1-form gauging on maximal branches and proposes the magnetic-quiver extension. Section~\ref{sec:conclusion} contains the conclusions and outlook, while the appendices collect the monopole-dressing and auxiliary technical computations.
\section{Setup and notation}
\label{sec:setup}
This section collects the ingredients used throughout the paper. 
First, the conventions for unitary linear CSM quivers and their Type IIB realisation are detailed in Section~\ref{subsec:UnitaryCSM_setup}.
Thereafter, Section~\ref{subsec:1FormSymm_tHooftSelfAnom} presents the concepts of 1-form symmetry and 't~Hooft self-anomaly.
Finally, Section~\ref{subsec:GlobalForma_2Groups_tHooftAnom} reviews the notions of global form of the 0-form symmetry group, the 2-group symmetry, and the mixed 't~Hooft anomalies.

\subsection{Unitary linear CSM quivers and brane systems}
\label{subsec:UnitaryCSM_setup}

\paragraph{Setup.}
Consider the 3d $\Ncal=3$ or $\Ncal=4$ supersymmetric QFTs realised in Type IIB string theory on the worldvolume of D3 branes ($-$) stretched between NS5 branes (~$\textcolor{red}{|}$~) and $(p_i,q_i)$ 5-branes (~\tikz[baseline=-0.2ex]{\draw[dash pattern=on 1.5pt off 1.5pt,blue,line width=.7pt] (-0.1,-0.1) -- (0.1,0.25);}~,~\tikz[baseline=-0.2ex]{\draw[dash pattern=on 1.5pt off 1.5pt,Mulberry,line width=.7pt] (-0.1,-0.1) -- (0.1,0.25);}~, $\dots$), with $p_i,q_i\in \Z$, see also~\cite{Hanany:1996ie,Gauntlett:1997pk,Gauntlett:1997cv,Kitao:1998mf,Bergman:1999na}. Conventions are such that $(\pm1,0)$ denotes NS5, while $(0,\pm1)$ denote D5 branes.
For generic $p_i$ and $q_i$, the resulting theory has no known Lagrangian description.
The theories realised by brane configurations with $p_i=\pm1$ (for all $i$), instead, are Lagrangian 3d unitary linear Chern--Simons Matter (CSM) theories. Without loss of generality\footnote{As discussed in \cite{Assel:2017eun}, one can always shift all $\kappa_i$ simultaneously by an integer such that one can achieve non-negative values throughout. This does not affect the D3 world-volume theory.}, one can restrict to $(p_i,q_i)=(1,\kappa_i)$ with non-negative $\kappa_i$.
The amount of supersymmetry they possess is in general $\Ncal=3$, and it enhances to $\Ncal=4$ if the brane setup involves only two types of 5-brane \cite{Assel:2022row}. By convention, these are chosen to be NS5 brane and a single type of $(1,\kappa)$ 5-brane.

In the $\Ncal=3$ case, the moduli space of the theory contains as many maximal branches as there are different $\kappa_i$, namely one maximal branch per 5-brane type \cite{Assel:2017eun,Marino:2025uub}.
In the $\Ncal=4$ case, since only NS5 and $(1,\kappa)$ 5-branes are present, the theory possesses two maximal branches, here denoted as A and B branches. This feature is a direct consequence of the R-symmetry, which factorises into $\surm(2)_\text{A}\times\surm(2)_\text{B}$ as in any other 3d $\Ncal=4$ theory.\\

A generic unitary CSM theory has gauge group $\Gcal=\prod_{i=1}^{L} \urm(N_i)_{k_i}$, where each factor comes with a CS-level $k_i\in\mathbb{Z}$. 
In the quiver notation employed throughout, each gauge group factor is displayed as a round node (~$\bigcirc$~) encoding the associated 3d $\Ncal=4$ vector multiplet. If the factor $\urm(N_i)_{k_i}$ comes with a non-zero CS-level, $k_i$ is written above the round node, now representing only the $\Ncal=2$ vector multiplet, and it is colour-coded with the 5-brane generating it.
Furthermore, each gauge group factor $\urm(N_i)_{k_i}$ comes with a topological $\urm(1)_{w_i}$ symmetry with fugacity $w_i$, together with a magnetic flux vector $m^{(i)}=(m^{(i)}_1,m^{(i)}_2,\dots,m^{(i)}_{N_i})\in\Z^{N_i}$.
A null flux vector is written as $(0,\dots,0)=\bullet$, while a string of zeros inside a flux vector is denoted with $\circ$: $(1,0,\dots,0)=(1,\circ)$.

As for the other fields, adjacent gauge nodes are connected by a $\urm(N_i)\times\urm(N_{i+1})$ bi-fundamental hypermultiplet, generically coloured in black. 
In the $\Ncal=4$ case, however, one can employ the axial symmetry (the $\urm(1)_R$ commutant inside the $\surm(2)_\text{A}\times\surm(2)_\text{B}$ R-symmetry) to distinguish between twisted and non-twisted bi-fundamental hypermultiplets, respectively coloured in blue and red.
An example of CSM quiver looks as follows.
\begin{align}
    \label{fig:example_quiver}
    \raisebox{-.5\height}{\includegraphics{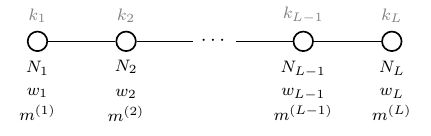}}
\end{align}

Finally, in the study of maximal branches, magnetic quivers are employed \cite{Cabrera:2018jxt,Cabrera:2019izd}: these are auxiliary non-CS 3d $\Ncal=4$ quiver theories that are read off from maximal branch phases of Type II intersecting brane configurations.
When writing them as quivers, square nodes (~$\Box$~) are used to encode flavour symmetries, while legs with an arrow represent non-simply laced edges.

\paragraph{Brane phases.}
Analogously to the NS5-D5 case \cite{Hanany:1996ie}, $(1,\kappa_i)$ and $(1,\kappa_j)$ 5-branes (with $i \neq j$) can be exchanged accounting for D3-brane creation/annihilation, also known as the Giveon--Kutasov (GK) duality \cite{Giveon:2008zn}.
In particular, adding two generic $(1,l)$ and $(1,r)$ 5-branes on the edges acting as spectators, one has the following.
\begin{align}
    \label{fig:GK_duality}
    \raisebox{-.5\height}{\includegraphics{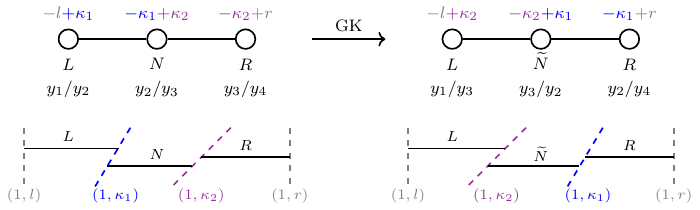}}
\end{align}
The final number of D3 branes in between is $\widetilde{N}=L+R-N+\abs{\det{\begin{smallmatrix} 1 & 1 \\ \kappa_1 & \kappa_2 \end{smallmatrix}}}$. Below each gauge rank $N_i$, the corresponding topological fugacities are written as ratios $w_i=y_i/y_{i+1}$ to emphasise the map.\\

The GK duality allows one to reach different brane phases for a given CSM theory. In particular, two phases are relevant for this paper:
\begin{itemize}
    \item \underline{Maximal branch phase}: Focusing on the $(1,\kappa_i)$ branch for a given $\kappa_i\in\N$, both the leftmost and rightmost 5-branes are $(1,\kappa_i)$, and there are no D3 moduli free to move along any other maximal branch.
    Moreover, the number of D3 segments ``frozen'' between adjacent $(1,\kappa_i)$ and $(1,\kappa_j)$ 5-branes is at most $\abs{\kappa_j-\kappa_i}$; hence, one can swap with the GK duality a $(1,\kappa_i)$ and a $(1,\kappa_j)$ 5-brane at most once without creating an excess amount of frozen D3s. 
    This construction imposes that
    \begin{equation}
        \sum_{i} k_i = 0 
        \qquad
        \text{ holds in a maximal branch phase}
        \,.
        \label{eq:basic_frac_monop_cond_maxbranch_1}
    \end{equation}
    \item \underline{Factorised phase}: For each $i$, all the $(1,\kappa_i)$ 5-branes are brought close together, so that only the node between each 5-branes' group has a non-zero CS-level, and the global symmetry factorisation is manifest.
    In particular, the $\Ncal=4$ case is the CSM-equivalent of the $T_\rho^\sigma[\surm(N)]$ phase of \cite{Gaiotto:2008ak, Cremonesi:2014uva}, where the symmetries of the theory are encoded in the partitions $\rho$ and $\sigma$ of the integer $N$ (see Appendix~\ref{app:CSM_TSUN} for details).
\end{itemize}
As an example, consider the following $\Ncal=4$ CSM setup.
\begin{align}
    \label{fig:example_phases}
    \raisebox{-.5\height}{\includegraphics{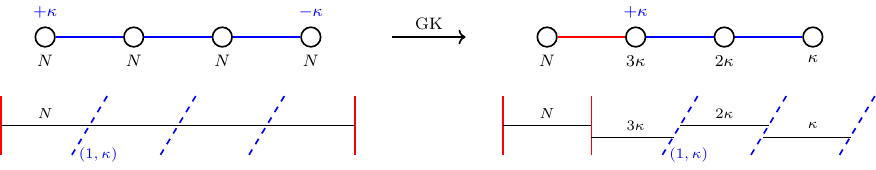}}
\end{align}
The A branch phase is represented on the left, while the factorised phase is drawn on the right; the two are connected by an iterative application of the GK duality.

\subsection{1-form symmetry and 't~Hooft self-anomaly}
\label{subsec:1FormSymm_tHooftSelfAnom}
For 3d $\Ncal=3$ and $\Ncal=4$ CSM theories the 1-form symmetry is determined from the analysis of charged lines and the conditions under which they can end on local operators. Only discrete 1-form symmetry groups $\Gamma^{(1)}=\mathbb{Z}_{g}^{(1)}$ are considered, due to the presence of the CS-interaction.
In three dimensional theories, both the generators and the charged objects for the 1-form symmetry are realised by line operators. The first ones are topological lines sourcing around them the $\Z_g$-valued background 2-cocycle $B_2$ that specifies the 1-form symmetry and the second ones are charged Wilson lines, not necessarily topological, on which the generators act by linking. 

A 1-form self-anomaly arises when the generating line also carries a non-trivial charge under the 1-form symmetry itself. 
The anomaly theory is, in general, given by the following 4d action:
\begin{equation}
\label{eq:intro_1-form_self}
    S_{\text{self}}^{\text{1-form}} \sim 2\pi\im  \int_{\mathcal{M}_4} \frac{\#}{g} B_2 \cup B_2 \, ,
\end{equation}
where $\#$ indicates the dependence on the other parameters of the theory.
It is determined by the charges under the 1-form symmetry group of non-gauge invariant operators sitting at the end of the lines. Consider the example in \eqref{fig:example_phases}: a 1-form symmetry line can end on a fractional flux monopole whose gauge charge is $q$ under the diagonal $\urm(1)$ contained in the $\urm(N_i)$ gauge nodes with CS-level. Then, the charge under the 1-form symmetry $\Z_{g}^{(1)}$ is given by $(q\,\text{mod}\,g)$, \cite{Bhardwaj:2022dyt}. 

The argument presented in this paper to derive the anomaly \eqref{eq:intro_1-form_self} focuses on the dressing of monopoles, reproducing the result of the charge computation. Since the anomaly conveys the obstruction to gauge the 1-form symmetry, the definition of $\#$ comes from the conditions under which the local operators, sitting at end of the generating lines, can be made gauge invariant after gauging $\mathbb{Z}_{g}^{(1)}$. See Section~\ref{subsubsec:fractional_monopoles} for details.

\subsection{Global forms, 2-groups, and 't~Hooft anomalies}
\label{subsec:GlobalForma_2Groups_tHooftAnom}
In this section the concepts necessary for the analysis of 2-group structures and mixed 't~Hooft anomalies are established, see \cite{Mekareeya:2022spm,Bhardwaj:2022dyt,Bhardwaj:2023zix} for further details. 
\paragraph{Notation.} 
To begin with, the notation is introduced.
\begin{itemize}
    \item Denote the gauge group of the UV theory by $\mathcal{G} =\prod_{i=1}^{L} \urm(N_i)$.
    \item The 1-form symmetry group is $\Gamma^{(1)} \subset \mathcal{Z}(\mathcal{G})$, with $\mathcal{Z}(\mathcal{G})$ the centre of $\mathcal{G}$.
    \item The faithfully acting 0-form symmetry group is $\mathcal{F}$: genuine local operators\footnote{Recall the conventions: a local operator is called \emph{genuine}, if it is gauge invariant without the need of extended operators attached to it. Hence, \emph{non-genuine} operators come with extended objects.} transform in representations of $\mathcal{F}$.
    \item The connected cover of the faithful symmetry group $\Fcal$ is denoted $F$: non-genuine local operators that can be used to end line defects can transform in representations of $F$ that are non-allowed representations of $\mathcal{F}$. Moreover, $\mathcal{F} = F/\mathcal{Z}$, where $\mathcal{Z} \subset \mathcal{Z}(F)$ is a subgroup of the centre of $F$.
    \item The 0-form symmetry is uniquely specified by the Lie algebra $\mathfrak{f} = \bigoplus_i \mathfrak{f}_i$, where each $\mathfrak{f}_i$ is either a $\mathfrak{u}(1)$ factor or a simple Lie algebra. $F$ and $\mathcal{F}$ are different Lie groups associated to the same Lie algebra $\mathfrak{f}$. This is referred to as different global forms.
    \item The second Stiefel--Whitney class is denoted as $\omega_2$: it measures the obstruction to lift the $\mathcal{F}$-bundle to an $F$-bundle. $\omega_2$ is a cohomology class in $H^2(B\mathcal{F}, \mathcal{Z})$, where $B\mathcal{F}$ is the classifying topological space of $\mathcal{F}$. For the case $\Fcal = \urm(1)$, the obstruction class to lift $\Fcal$ to its $n$-fold cover $F=\widetilde{\urm(1)}$ is defined as $\omega_2= (c_1\,\text{mod}\,n)$, using the first Chern class of $\urm(1)$ bundles.
\end{itemize}

\subsubsection{2-group symmetry}
\label{subsubsec:2group_intro}
In the presence of a 0-form symmetry group $\mathcal{F}$ and 1-form symmetry group $\Gamma^{(1)}$, a theory experiences a 2-group symmetry if there is an interplay between them \cite{Benini:2018reh,Cordova:2018cvg}.
The theory, then, can be coupled to a 2-group background field that, in general, consists of a 1-form  background $A_1$ for the group $\mathcal{F}$, a 2-form $B_2$ for the group $\Gamma^{(1)}$, and a rule for their gauge transformations determined by two other variables: the action $\rho$ of the 0-form symmetry on the 1-form and the Postnikov class $\beta$.
The first is given by a group homomorphism $\rho: \mathcal{F} \rightarrow \text{Aut}(\Gamma^{(1)})$.
The second is a group cohomology class $\beta \in H^3(B\mathcal{F}, \Gamma^{(1)})$ that encodes the non-associativity of the 0-form symmetry; namely, it is a function 
\begin{equation}
    \beta: \mathcal{F}\times \mathcal{F} \times \mathcal{F} \rightarrow \Gamma^{(1)} 
\end{equation} 
implying the presence of a 1-form symmetry defect in the junction of three 0-form symmetry operators. This class parametrises how the two symmetries can mix, namely the background for the discrete 1-form symmetry is not closed but is fixed by
\begin{equation}
\label{eq:setup_1-form_background}
    \delta B_2 = A_1^{\, \,*} \beta \,,
\end{equation}
where $A_1^{\, \,*}$ denotes the pull-back on spacetime of the Postnikov class via the background connection $A_1$ of $\mathcal{F}$, seen as a map from spacetime to $B\mathcal{F}$.

A \emph{non-trivial} 2-group structure is characterised by the short exact sequence
\begin{equation}
    \label{shortexact}
    0 \rightarrow \Gamma^{(1)} \rightarrow \mathcal{E} \rightarrow \mathcal{Z} \rightarrow 0 \, ,
\end{equation}
in which $\mathcal{E}$ is given by a non-trivial extension of $\mathcal{Z}$ by $\Gamma^{(1)}$, so the short exact sequence is non-split. 
This allows one to understand how the Postnikov class $\beta$ is built from the second Stiefel--Whitney class $\omega_2$:
\begin{equation}
    \beta = \text{Bock}(\omega_2) \, .
\end{equation}  
The connecting homomorphism Bock 
\begin{equation}
    \text{Bock}: H^2(B \mathcal{F}, \mathcal{Z}) \rightarrow H^3(B \mathcal{F}, \Gamma^{(1)})
\end{equation}
belongs to the long exact sequence in cohomology associated to \eqref{shortexact}. 
Moreover, the 2-group is called \emph{trivial} if its defining short exact sequence splits, namely if there exists an isomorphism 
\begin{equation}
    \Gamma^{(1)} \times \mathcal{Z} \,  \xrightarrow{\cong} \, \mathcal{E} \, .
\end{equation}

In the following, the non-trivial 2-groups considered have trivial action $\rho$ and non-trivial Postnikov class. For a generalisation of this analysis, including a non-trivial action, see Section \ref{subsubsec:charge_conj}: an example of a non-trivial $\rho$ is realised considering the charge conjugation symmetry, see also \cite{Benini:2018reh}. As a remark, in the examples in which the only 0-form global symmetry of the theory is given by a $\urm(1)$ abelian factor, the 2-group with a discrete 1-form symmetry $\Gamma^{(1)}$ always has a trivial Postnikov class, giving rise to a trivial 2-group. In fact, since the Chern class $c_1$ of $\urm(1)$ bundles (that is a 2-form valued in $\Z$) is the generator of the group $H^2(B \, \urm(1), \Z) \cong \Z$, it follows that $H^{2n+1}(B \, \urm(1), \Z) \cong 0$ for every $n\in \N$. This implies that the map Bock needs to be trivial, see also \cite[Sec. 6.2]{Benini:2018reh}.

The main technique that is applied to detect the 2-group structure is the study of 1-form symmetry defects. The lines that are charged under $\Gamma^{(1)}$ have a finite $g$-fold power that can end on some dressed monopole operators of the CSM quivers. These monopoles transform non-trivially under the 0-form symmetry. If they are charged under the centre $\mathcal{Z}$, they are not in a representation of $\mathcal{F}$ and this implies the presence of a non-trivial 2-group symmetry.

\subsubsection{Mixed 't~Hooft anomalies}
In order to discuss mixed 't~Hooft anomalies between 1-form and 0-form symmetries, it is necessary to distinguish between the cases of trivial and non-trivial 2-group structure. For the non-trivial case, it is possible to study a mixed anomaly between a 2-group background $\widetilde{B}_2$ and some 0-form symmetry group $H$ that does not participate in the 2-group. The background $\widetilde{B}_2$ is valued in $\mathcal{E}$ and it is a combination of the 1-form symmetry 2-cochain $B_2$ and of the obstruction class $\omega_2$ of the 0-form symmetry $\mathcal{F}$,
\begin{equation}
\label{eq:2-groupbackground}
    \widetilde{B}_2 = i(B_2 ) + \tilde{\omega}_2 \, .
\end{equation}
In order to obtain (\ref{eq:2-groupbackground}) one needs to apply to $B_2$ the inclusion map $i: \Gamma^{(1)} \rightarrow \mathcal{E}$ and to $\omega_2$ the lift to $\mathcal{E}$ under the projection map $\pi: \mathcal{E} \rightarrow \mathcal{Z}$. The two maps $i$ and $\pi$ are the ones that define the short exact sequence \eqref{shortexact}.
An anomaly with the general structure\footnote{For a more formal definition of the anomaly actions see \cite{Bhardwaj:2022dyt,Bhardwaj:2023zix}.}
\begin{equation}
    S_{\text{2-group mixed}} \sim 2\pi\im  \int_{\mathcal{M}_4} \# \, \widetilde{B}_2 \cup \omega_2^H \, ,
\end{equation}
depending on the degree two characteristic class $\omega_2^H$ of the 0-form symmetry bundle for the group $H$, is examined in the examples, for the particular case of $H=\urm(1)$. 

In contrast, if the 2-group is trivial one can examine the mixed anomaly between the 1-form and 0-form symmetries. The presence of an anomaly is determined from the local operators of the theory that can be used to end the line defects generating the 1-form symmetry. If such non-genuine local operators transform in a representation of $F$ which is not an allowed representation of $\mathcal{F}$, then it motivates the following anomaly theory:
\begin{equation}
    S_{\text{mixed}} \sim 2\pi\im  \int_{\mathcal{M}_4} \# \, B_2 \cup \omega_2 \, .
\end{equation}
The anomaly coefficient $\#$ depends on the symmetries; for example if $B_2$ is a $\Z_{g}$-valued background and $\omega_2$ is a $\Z_N$-valued characteristic class, then $\#$ is fixed to be proportional to $1/\gcd(N,g)$, because of the definition of the cup product, see \cite{Bertolini:2026odb}.
The information encoded in this action is the obstruction to gauge the 1-form symmetry $\Gamma^{(1)}$ and the 0-form symmetry $\mathcal{F}$ at the same time, which is consistent with the perspective of line defects. In fact, after gauging the 1-form symmetry a fractional monopole that ends the line generator becomes a genuine local operator. If the monopole is charged under $\mathcal{Z}$, then the actual 0-form symmetry group of the theory after gauging is larger than $\mathcal{F}$; this prevents gauging the 0-form symmetry with global form $\mathcal{F}$. It is only possible to gauge the 0-form symmetry via the group F.

Note that this reasoning is similar (but not identical) to the argument presented in the 2-group case. Here the lines under consideration are the generating defects of the 1-form symmetry which can end only on fractional flux monopole operators. In the 2-group case, instead, the 1-form symmetry is still a true global symmetry of the system, so only the $g$-fold power of the charged line can end on integer flux monopoles.

Consider finally the mixed 't~Hooft anomalies between 0-form symmetries factors. Suppose $\Fcal= \Fcal_A \times \Fcal_B$ and the connected cover $F=F_A \times F_B$, for example in $\Ncal=4$ theories. The general structure of the anomaly action is given by 
\begin{equation}
\label{eq:0-form_anomaly}
    S_{0\text{-form}} \sim 2\pi\im  \int_{\mathcal{M}_4} \# \,  \omega^A_2 \cup \omega^B_2 \, ,
\end{equation}
where $\omega_2^{A,B}$ are the obstruction classes to lift $\Fcal_{A,B}$ to $F_{A,B}$, respectively. One possible strategy to determine the coefficient of the anomaly is similar to the 1-form symmetry self-anomaly discussion. In all the examples the 0-form symmetries are topological symmetries, whose background flux is induced by the insertion of a Wilson line in determinant representation of some gauge node. The anomaly coefficient in \eqref{eq:0-form_anomaly} is given by the charge of the monopole operator ending the Wilson line inducing the background for $F_A$, under the centre of $F_B$. The same analysis can be repeated for 0-form self-anomalies, computing the charge under the same symmetry.
\section{\texorpdfstring{Symmetries and anomalies of 3d $\Ncal\geq3$ CSM theories}{Symmetries and anomalies of 3d CSM theories}}
\label{sec:symm_anom}
In this section the symmetry structures of 3d CSM theories with $\Ncal =3$ or $\Ncal=4$ supersymmetry are analysed. The investigation of 1-form symmetries, 2-groups and global forms of 0-form symmetries is detailed, together with the study of 't~Hooft anomalies. 

\subsection{1-form symmetry}
\label{subsec:1-form_N=3,4}
Consider a Lagrangian $\Ncal\geq 3$ CSM linear quiver of the general form \eqref{fig:example_quiver}, with gauge group $\Gcal=\prod_{i=1}^L \urm(N_i)_{k_i}$: it is the world-volume theory on the D3 branes in between $L+1$ many $(1,\kappa_i)$ 5-branes\footnote{Recall from \cite{Assel:2017eun}, one can always shift all $\kappa_i$ simultaneously by an integer to set, say, $\kappa_1=0$, such that there is always at least on NS5 brane present.} (ordered from left to right), for $i=1,\ldots,L+1$ and $\kappa_i \in \N$, such that the CS-levels are $k_i = \kappa_{i+1}-\kappa_i$, for $i=1,\ldots,L$. Throughout, \emph{assume} that the CSM theory has no fundamental flavours (\ie no $\text{D5}=(0,1)$ branes). The reason for the assumption is clarified below.

\paragraph{From lines.}
To deduce the 1-form symmetry, consider Wilson lines as the charged objects. 
Firstly, the fundamental Wilson lines $W_{1\otimes \ldots \otimes 1\otimes F\otimes 1 \ldots \otimes 1}$ of some $\urm(N_i)_{k_i}$ cannot end on any local operator.
To see this, recall that all hypermultiplets transform bi-fundamentally; hence, they are insufficient to compensate the gauge charges of the Wilson line. Moreover, all fundamental monopole operators of a fixed gauge node transform either in some $k_i$-fold tensor products or trivially if $k_i=0$. 
However, the $|k_i|$-th tensor power of such a fundamental Wilson line can end on a local operator; specifically, on the (anti-)unit monopole of $\urm(N_i)_{k_i}$. 
Secondly, all fundamental Wilson lines are equivalent via the bifundamental hypermultiplet in between the two adjacent gauge nodes. This equivalence of all Wilson lines and the equivalence of the $k_i$-fold tensor powers to a trivial Wilson line implies
    \begin{align}
        (W_{F\otimes 1 \ldots \otimes 1})^{\otimes |k_i|} \sim 1 \,\,\,\,\, \forall i\in\{1,\ldots,L\}
        \quad 
        \Longrightarrow 
        \quad 
        (W_{F\otimes 1 \ldots \otimes 1})^{\otimes g}
        \sim 1
    \end{align} 
    by Bézout’s lemma. Therefore, the 1-form symmetry group is
    \begin{align}
        \Z_g^{(1)} \qquad\text{with}\qquad g=\gcd(k_1,k_2,\ldots) = \gcd(\kappa_1,\kappa_2,\ldots) \;. 
        \label{eq:gcd_N=3_QFT}
    \end{align}

\paragraph{From branes.}
In a 3d $\Ncal=4$ NS5-D5-D3 brane setup, Wilson lines are realised by F1 strings, while vortex lines are realised by D1 branes~\cite{Assel:2015oxa}.
One expects general Wilson-vortex lines to be realised as $(p,q)$ strings \cite{Schwarz:1995dk,Simon:2011rw,Griguolo:2021rke,Hayashi:2025guk,BrolisMarinoPasquetti:WIP}.
Fundamental strings can end on D5 branes, while D1 can end on NS5 branes. This means that any Wilson line can end, specifically on some D5 brane, which is the brane equivalent of the field theory statement that electric Wilson lines are screened by fundamental flavours.
Likewise, the D1 can end on NS5 branes, therefore there is no 1-form symmetry as vortex lines can end on some monopole operators.
In terms of the $\Z^2$ lattice of $(p,q)$-strings, one computes the set of lines that cannot end simply by
\begin{align}
    \Z^2 \slash \langle (1,0), (0,1) \rangle \cong \{0\} \,,
\end{align}
where the quotient has been taken by the D1 $=(1,0)$ brane and F1 $=(0,1)$ string, as they can end. As anticipated, the quotient is trivial, namely there is no 1-form symmetry.

On the other hand, in NS5-$(1,\kappa_i)$-D3 brane systems realising 3d $\Ncal\geq 3$ CSM theories, the fundamental string cannot end as there are no D5 branes.
One can ask which $(p,q)$ strings can end on the available 5-branes.
The known answer is D1 and $(1,\kappa_i)$ lines.
One constructs the set of lines that cannot end by taking the quotient of the $(p,q)$ lattice by the lines that can end:
\begin{align}
    \Z^2 \slash \langle (1,0) ,(1,\kappa_2),\ldots,(1,\kappa_L) \rangle \cong \Z_g  \,.
\end{align}
This gives a $\Z_g^{(1)}$ 1-form symmetry, in particular saying that the F1-string cannot end but only $\mathrm{F1}^{\otimes g}$ can, since it can be written as a combination of the endable lines. 

As a remark, if the brane configuration includes also D5 branes (\ie fundamental flavours), then the $(p,q)$ lattice quotient trivialises, and there is no 1-form symmetry.
Therefore the absence of D5 branes is assumed throughout.

\subsubsection{Central Gukov--Witten defect}
\label{sec:Gukov--Witten}
The $\Ncal\geq 3$ linear CSM quiver~\eqref{fig:example_quiver} has gauge group $\mathcal{G}=\prod_{r=1}^L \urm(N_r)_{k_r}$ and 1-form symmetry $\Z_g^{(1)}$. The generator $D$ of $\Z_g^{(1)}$ can be realised via a Gukov--Witten defect \cite{Gukov:2006jk,Gukov:2008sn} in the UV Lagrangian gauge theory. Denote the $\urm(N_r)$ gauge field by $A^{(r)}$, for any $r\in\{1,\ldots,L\}$. 
A fractional central Gukov--Witten type defect is defined as follows: around a small circle linking the line of $D$, one imposes 
\begin{subequations}
    \label{eq:diag_centre_defect}
    \begin{align}
       \frac{1}{2 \pi} \oint A^{(r)} = \frac{1}{g}\mathds{1}_{N_r} 
       \quad \text{for all }  r.
    \end{align}
    Equivalently, near the line, one has 
    \begin{align}
    A^{(r)}\sim \frac{1}{g} \mathds{1}_{N_r} \diff \theta 
    \quad \text{for all }  r\,,
    \end{align}
    with $\theta$ the angular coordinate.
\end{subequations}
 This monodromy is specified by 
 \begin{align}
  \alpha_D=\frac{1}{g}(\mathds{1}_{N_1},\mathds{1}_{N_2},\ldots,\mathds{1}_{N_L})   \;,
 \end{align}
which is a cocharacter of $\mathcal{G}/\Z_g$ and not of $\mathcal{G}$. 

There are some properties of $D$ worth mentioning. Firstly, the braiding between $D$ and a fundamental Wilson line $W_r$ at the $r$-th gauge group factor is computed via 
\begin{align}
    M(D,W_r) = \exp(2 \pi \im\; \alpha_D \, \widetilde{\mu}_r)
        = \exp(2 \pi \im\; \frac{1}{g}(1,\ldots,1) \cdot \mu_r)
        = \exp(\frac{2\pi \im}{g}) \,,
\end{align}
wherein $\mu_r \cong(1,0,\ldots,0)$ denotes the fundamental weight of $\urm(N_r)$ (understood as weight $\widetilde{\mu}_r$ of the full group $\mathcal{G}$) and 
$\alpha_D \, \widetilde{\mu}_r$ denotes the dual pairing. Hence, all fundamental Wilson lines acquire the same phase under the 1-form symmetry.

Secondly, due to the CS-interactions, $D$ is charged under $\mathcal{G}$ and these charges are evaluated via $q_D= K_{\mathrm{UV}}\,\alpha_D$ where $K_{\mathrm{UV}} = \mathrm{diag}(k_1 \mathds{1}_{N_1},k_2\mathds{1}_{N_2},\ldots,k_L \mathds{1}_{N_L})$ denotes the matrix of CS-terms. Thus, the charges $q_D$ restricted to $\urm(N_j)$ are computed to be
\begin{align}
    (q_D)_j =  \frac{k_j}{g} (1,\ldots,1) \,.
\end{align}
Therefore, due to the presence of the CS-terms, $D$ is dyonic in nature and transforms in the $\mathrm{det}^{k_j/g}$ representation of $\urm(N_j)_{k_j}$ and transforms trivially under any $\urm(N_i)_{k_i=0}$ with vanishing CS-level.

Thirdly, the CS-interaction is the source of a possible 1-form self-anomaly. The reason is that $q_D$ are charges under $\mathcal{G}$, but $\Gamma^{(1)}\subset \mathcal{Z}(\mathcal{G})$. Computing the charges of $D$ under $\urm(1)_{\mathrm{diag}}\subset \mathcal{G}$, leads to 
\begin{align}
 p= \tr{q_D}= \sum_{r=1}^L \frac{k_r}{g}  N_r    \,.
\end{align}
Thus, $D$ has $\Z_g^{(1)}\subset \urm(1)_{\mathrm{diag}}$ charge
\begin{align}
    p \bmod g = \sum_{r=1}^L \frac{k_r}{g}  N_r \bmod g \,
    \label{eq:1-form_self-anomaly_GW}
\end{align}
and whenever this is non-trivial, there is a 1-form self-anomaly. 
This is also captured by topological spin\footnote{Throughout, the self-anomaly of a $\Z_n^{(1)}$ symmetry generated by an abelian line $u$ is labelled by the integer $p$ defined through the topological spin $h[u]=\frac{p}{2n}$; see for instance~\cite{Hsin:2018vcg}.}
\begin{align}
    h[D] = \frac{1}{2} \alpha_D^T \, K_{\mathrm{UV}} \, \alpha_D  = \frac{1}{2 g}  \sum_{r=1}^L \frac{k_r}{g}  N_r
    = \frac{p}{2g}\,,
\end{align}
which matches the $p$ expression \eqref{eq:1-form_self-anomaly_GW}.

\subsubsection{Fractional monopole operators}
\label{subsubsec:fractional_monopoles}
Given that the generating line $D$ is specified by a fractional cocharacter $\alpha_D$, one can also think about the non-genuine monopole operators $\mono{\text{frac}}$ with fractional magnetic flux at the end point of $D$, understood as vortex line. This offers a parallel perspective to determine the 1-form self-anomaly. If the non-genuine monopole operator $\mono{\text{frac}}$ cannot be dressed by the matter fields in the CSM quiver, then this end point can never become a genuine monopole operator if one attempts to gauge $\Z_g^{(1)}$. This failure is a manifestation of the 1-form self-anomaly~\eqref{eq:intro_1-form_self}.

Concretely, for a linear $\Ncal\geq3$ CSM quiver~\eqref{fig:example_quiver},
the 1-form self-anomaly is derived by inspecting the failure of a bare non-genuine monopole operator $\mono{\text{frac}}$ with fractional magnetic flux to be dressable by matter fields into a gauge-invariant operator.
Suppose one wants to gauge $\Gamma^{(1)}=\Z_g^{(1)}$, then the CSM quiver gauge group changes from $\Gcal$ to $\Gcal/\Z_g^{(1)}$ and $\alpha_D$ is a legitimate cocharacter of the new gauge group. While the coweight lattice of $\Gcal$ is $\Z^{\mathrm{rk}\Gcal}$, the magnetic lattice of $\Gcal\slash \Z_g^{(1)}$ is $\bigcup_{a=0}^{g-1} (\Z^{\mathrm{rk}\Gcal} + a\, \alpha_D)$, see for instance \cite{Bourget:2020xdz}. In practice, the magnetic fluxes $m^{(i)}$ are \emph{uniformly} shifted by $\frac{a}{g}$, for some $a\in \{0,1,\ldots,g-1\}$. 
For a bare monopole with such fractional flux, one can at most use integer powers of the bifundamental matter fields to compensate the gauge charges induced by the CS-terms. Therefore, there exists a charge cancellation condition at every gauge node, see Appendix~\ref{app:dressed_monopoles} for an example. Combining the conditions from all nodes implies a single charge cancellation condition for the $\urm(1)_{\mathrm{diag}}\subset \Gcal$, which reads
\begin{align}
    \sum_{i=1}^L \left(  k_i  \sum_{j=1}^{N_i} m_j^{(i)}\right) + 
    \frac{a}{g} \sum_{i=1}^L N_i k_i  =0  
    \quad \Longleftrightarrow \quad
    \sum_{i=1}^L \left(  \tilde{k}_i  \sum_{j=1}^{N_i} m_j^{(i)}\right) + 
    a\frac{\sum_{i=1}^L N_i \tilde{k}_i}{g}  
    =0 
\end{align}
with $g=\gcd(k_1,k_2,\ldots,k_L)$, $k_i \equiv g \cdot \tilde{k}_i$, and $a\in\{0,1,\ldots,g-1\}$.
It follows that an integer solution for all $\{m_j^{(i)}\}$ exists only if 
\begin{align}
   g \mid (a \cdot p) \,, 
    \qquad \text{with}\quad 
     p \coloneqq \sum_{i=1}^L N_i \tilde{k}_i \equiv \sum_{i=1}^L N_i\frac{k_i}{g}
    \,.
\label{eq:Unitary_anomaly_fractional}
\end{align}
The interpretation is as follows:
\begin{itemize}
    \item If $g\mid p $, then for every $a$ the fractional monopole can be dressed. Hence, the $\Z_g^{(1)}$ 1-form symmetry is anomaly-free and can be consistently gauged.
    \item If $g\nmid p  $, but $\exists$ $a\in\{0,1,\ldots,g-1\}$ such that $g\mid p\cdot a $, then there exists a subgroup\footnote{It is a subgroup because $a=0$ solves the constraint automatically and gives rise to the identity element.} 
    \begin{align}
    \label{eq:anomaly-free_1-form}
    \Z_h^{(1)}\subset \Z_g^{(1)} \qquad \text{with} \quad h=\gcd(g,p) \, ,
    \end{align} 
    (generated by all such $a$'s) that is anomaly-free and can be gauged. This is manifest as only for certain $a$'s the fractional monopole enters the physical spectrum.
    The claim is that this scenario can only occur if $h >1$. To see this, if $h=1$ then $g \mid p \cdot a$ is equivalent to $g \mid a$ by Euclid's lemma, which can never be true for any $a\in\{0,1,\ldots,g-1\}$. For $h>1$, redefine $g \equiv h \cdot \tilde{g}$ and $p \equiv h \cdot \widetilde{p}$ such that $\gcd(\widetilde{p},\tilde{g})=1$. Then the condition becomes
    \begin{align}
        g \mid p \cdot a 
        \quad \Leftrightarrow\quad 
        \tilde{g} \mid \widetilde{p} \cdot a
        \quad \xRightarrow{\text{Euclid}}\quad
        \tilde{g} \mid a
    \end{align}
    which is always true for some $a\in\{0,1,\ldots,g-1\}$, because $0<\tilde{g}<g$ and $\tilde{g}\in \N$. The \emph{maximal anomaly-free subgroup} is hence given by $\Z_h^{(1)}$.
\end{itemize}
Therefore, the condition \eqref{eq:Unitary_anomaly_fractional} is understood as the 1-form anomaly 
\begin{align}
    S_{\text{self}}^{\text{1-form}} \sim 2\pi\im  \int_{\mathcal{M}_4} \frac{p}{g} B_2 \cup B_2
    \,,
\end{align}
this matches \eqref{eq:1-form_self-anomaly_GW}, because $p \bmod g$ is the $\Z_g^{(1)}$ charge of $D$.
Note that the anomaly is invariant under GK dualities.
In fact, the contribution to the self-anomaly label $p$ of the two dual theories can be deduce using \eqref{fig:GK_duality}, and one observes that they agree $\bmod \, g$.

\subsection{\texorpdfstring{0-form and 2-group symmetries in $\Ncal=4$ CSM}{0-form and 2-group in N=4 CSM}}
\label{subsec:general_N=4}
This section collects criteria to identify the global form of 0-form symmetries, detect 2-group structures and determine 't~Hooft anomalies in 3d $\Ncal =4$ CSM theories.

\subsubsection{Global form of the 0-form symmetry group}
For a CSM brane system composed of $n_A$ NS5 and $n_B$ $(1,\kappa)$ 5-branes, one can always move to the brane phase in which all NS5 branes are on the left and all $(1,\kappa)$ 5-branes on the right, namely the factorised phase defined in Section~\ref{subsec:UnitaryCSM_setup}. Thereafter, the CSM quiver is of the following form. 
\begin{align}
    \label{eq:example_quiver_symmetries}
    \raisebox{-.5\height}{\includegraphics{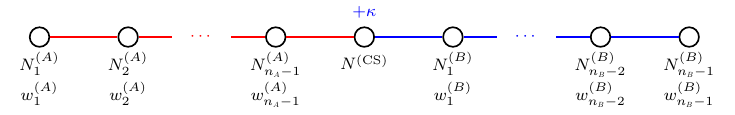}}
\end{align}
Note that the 1-form symmetry is $\Z_\kappa^{(1)}$. Due to the graph underlying the CSM quiver, most of the 0-form symmetry structures can be derived explicitly from it. 

\paragraph{Setup.}
Let $X=A,B$ and focus, to begin with, on the $X$ branch 0-form symmetry $\Fcal_X$. 
Denote by $\mathcal{B}_X$ the connected balanced components on side $X$ and by $\mathcal{O}_X$ the set of its overbalanced nodes. If $I \in \mathcal{B}_X$ contains $\ell_I$ balanced nodes, define $\mathfrak{f}_I = \surmL(\ell_I+1)$. The 0-form Lie algebra on side $X$ is 
\begin{align}
            \ffrak_X = \urmL(1)^{|\mathcal{O}_X|} \oplus
            \left[
            \bigoplus_{I \in \mathcal{B}_X} \surmL(\ell_I+1) 
            \right]
            \label{eq:def_ffrak_X}
\end{align}
which has compact covering group
\begin{align}
           F_X =\urm(1)^{|\mathcal{O}_X|} \times  \prod_{I \in \Bcal_X} \surm(\ell_I+1)  
           \quad \text{with} \quad
           Z_X = Z(F_X)
           \,,
\end{align}
where the centre group $Z_X$ and its Pontryagin dual are explicitly given by
\begin{align}
            Z_X = \urm(1)^{|\mathcal{O}_X|} \times  \prod_{I \in \mathcal{B}_X} \Z_{\ell_I+1}
            \; ,\qquad 
            \widehat{Z}_X \cong \Z^{|\mathcal{O}_X|} \oplus \bigoplus_{I \in \mathcal{B}_X} \Z_{\ell_I+1} \,.
\end{align}

\paragraph{Faithful symmetry on maximal branch operators.}
To determine the faithful symmetry group $\Fcal_X$, one proceeds similarly to \cite{Bhardwaj:2023zix}. For every $r\in\Ocal_X$, choose the unit monopole, denoted by $\mathfrak{M}_r$, and compute its $Z_X$ charges. To do so, a few subtleties are to be addressed:
    \begin{itemize}    
    \item The $X$ branch UV topological fugacities are $w_i^{(X)}$ for $i=1,2,\ldots,n_X-1$.
    
    \item For any $I\in\Bcal_X$, the UV fugacities $\{w_i^{(X)}\}_{i \in I}$ are mapped to IR fugacities $\{x_i\}_{i\in I}$ via the $\surmL(\ell_I+1)$ Cartan map $w_i^{(X)} = \prod_{j\in I} x_j^{C_{ij}}$ with $C_{ij}$ the $A_{\ell_I}$ Cartan matrix. 
    
    \item For any $r\in \Ocal_X$, one needs to pay attention to the fact that abelian factors might mix with non-abelian factors, see for instance~\cite{Nawata:2023rdx,Marino:2025uub}.
\end{itemize} 

Suppose an overbalanced node $r\in\Ocal_X$ is connected to a balanced subset $J\in\Bcal_X$ on the right, then there exist monopole operators that transform as $[1,0,\ldots,0]^1$ (respectively \ $[0,\ldots,0,1]^{-1}$) under the $\surmL(\ell_J+1)_X \oplus \urmL(1)_X$. The representation is denoted with the Dynkin label for $\surmL(\ell_J+1)_X$ and the superscript is the charge under $\urmL(1)_X$. To realise this, one needs the map:
\begin{subequations}
\label{eq:mix_sym_Fcal_N=4}
\begin{equation}
w_r^{(X)}=q_r/x_1 \, ,
\end{equation}
where $x_1$ is the fundamental weight fugacity of $\surmL(\ell_J+1)_X$.
Hence, the unit monopole $\mathfrak{M}_r$ has $\surmL(\ell_J+1)_X$ centre charge $1 \bmod  (\ell_J+1)$.

Suppose an overbalanced node $r\in\Ocal_X$ is connected to a balanced subset $I\in\Bcal_X$ on the left, then there exist monopole operators that transform as $[0,0,\ldots,1]^1$ (resp.\ $[1,\ldots,0,0]^{-1}$) under the $\surmL(\ell_I+1)_X \oplus \urmL(1)_X$. To realise this, one needs the map: 
\begin{equation}
w_r^{(X)}=q_r/x_{\ell_I} \, , 
\end{equation}
\end{subequations}
where $x_{\ell_I}$ is the anti-fundamental weight fugacity of $\surmL(\ell_I+1)_X$. 
Thus, the unit monopole $\mathfrak{M}_r$ has $\surmL(\ell_I+1)_X$ centre charge $\ell_I \bmod  (\ell_I+1) = -1\bmod  (\ell_I+1)$.

If there is an overbalanced node $r\in\Ocal_X$ that is connected to the left and right to two balanced subsets $I,J\in\Bcal_X$ of nodes, then both maps apply simultaneously, as there exists a collection of monopole operators that transform as $([0,\ldots,0,1]_{\surm(\ell_I+1)} \otimes [1,0,\ldots,0]_{\surm(\ell_J+1)})^{+1}$.
The unit monopole $\mathfrak{M}_r$ has $\surmL(\ell_I+1)_X \oplus \surmL(\ell_J+1)_X$ centre charges  $( -1\bmod  (\ell_I+1),1 \bmod  (\ell_J+1))$. 

Therefore, the complete centre characters for the unit monopole $\mathfrak{M}_r$ for any $r\in\Ocal_X$ are given by:
\begin{align}
    q_r^X = (e_r ; (c_{r,I})_{I\in \Bcal_X})
\end{align}
where $e_r$ is the $r$-th basis vector of $\Z^{|\mathcal{O}_X|}$ and the centre charge $c_{r,I}$ is, based on the above discussion, computed from the incidence $m_{r, j}$ of $r\in\Ocal_X$ with the Dynkin node $j$ in a balanced subset $I\in\Bcal_X$ via
\begin{align}
    c_{r,I}  = \sum_{j\in I} m_{r, j} j \bmod (\ell_I+1) \;.
\end{align}
Define the subgroup generated by genuine unit-monopole characters
\begin{align}
    \label{eq:def_Y_X}
    Y_X \coloneqq \mathrm{Span}_{\Z} \{q_r^X \; | \; r\in \Ocal_X \} \subset \widehat{Z}_X \;.
\end{align}
The faithful symmetry group is realised by the following quotient:
    \begin{equation}
            \begin{aligned}
    \widehat{\Zcal}_X \coloneqq \widehat{Z}_X \slash Y_X 
    \quad &\text{with Pontryagin dual} \quad 
    \Zcal_X \coloneqq (\widehat{\Zcal}_X)^\vee  \\
   \quad &\text{such that} \quad  \Fcal_X = F_X \slash \Zcal_X \;.
    \end{aligned}
    \end{equation}

\paragraph{CS-node probe and its extended group.}
The central node $\urm(N^{(\CS)})$ is the only one with CS-interactions; hence, the unit monopole $\mono{\mathrm{probe}}$ of that node is a non-genuine operator and defines a probe for the extended 0-form group $F$. This is a genuine new feature of 3d $\Ncal=4$ CSM theories compared to the 3d $\Ncal=4$ non-CS theories of \cite{Bhardwaj:2023zix,Nawata:2023rdx}. 
Define 
\begin{subequations}
\begin{alignat}{3}
    F_0 &\coloneqq F_A \times F_B  &
    \;, \qquad 
    Z_0 &\coloneqq Z_A \times Z_B  &
    \;, \qquad 
    \widehat{Z}_0 &= \widehat{Z}_A \oplus \widehat{Z}_B \;, \\
   &\text{and } &  Y_0 &\coloneqq Y_A \oplus Y_B  &
     \;, \qquad 
      \Zcal_0 &\coloneqq \Zcal_A \times \Zcal_B \;,
\end{alignat}
\end{subequations}
where $\Zcal_0$ is obtained as Pontryagin dual of $\widehat{Z}_0 \slash Y_0$.

The 0-form symmetry acting faithfully on genuine local operators is
\begin{align}
    \Fcal = F_0 \slash  \Zcal_0 = \Fcal_A \times \Fcal_B \,.
\end{align}
Denote by $p_{\mathrm{CS}} = (p_A,p_B) \in \widehat{Z}_0 $ the complete character of a dressed unit monopole $\mathfrak{M}_{\mathrm{probe}}$ at the CS node on which $(W_{1\otimes \ldots \otimes 1\otimes \overline{F}_{N^{(\text{CS})}} \otimes 1 \otimes \ldots \otimes 1})^{\otimes \kappa}$ can end\footnote{More precisely, the unit monopole sits at the end of the Wilson line $W_{\mathrm{Sym}^\kappa(\overline{\mathbf{F}}_{N^{(\text{CS})}})}$, but $\mathrm{Sym}^\kappa(\overline{\mathbf{F}}_{N^{(\text{CS})}})$ is realised as a highest weight module inside $\overline{\mathbf{F}}_{N^{(\text{CS})}}^{\otimes \kappa}$.}. The relevant information is encoded in the class $[p_{\mathrm{CS}}] \in \widehat{Z}_0 \slash Y_0$, because two dressings of the same endpoint can differ by genuine monopoles and, hence, by an element of $Y_0$.

The minimal connected group under which genuine operators and non-genuine endpoints are all honest representations is
\begin{equation}
    F= \frac{F_0}{\Zcal_{\text{tot}}}= \frac{F_A \times F_B}{\Zcal_{\text{tot}}} \, ,
\end{equation}
where $\Zcal_{\text{tot}}$ is defined by
\begin{equation}
    Y_{\mathrm{tot}}= Y_0 + \mathrm{Span}_\Z\{p_{\mathrm{CS}}\}\;,\quad
    \widehat{\Zcal}_{\text{tot}}= \widehat{Z}_0 \slash Y_{\mathrm{tot}} 
    \;, \text{ with Pontryagin dual } 
    \Zcal_{\text{tot}} = ( \widehat{\Zcal}_{\text{tot}})^\vee \, .
\end{equation}
To determine $p_{\mathrm{CS}}$, explicit computations led to the following four distinct cases\footnote{All entries of $p_{\mathrm{CS}}$ mean classes in $\widehat{Z}_0\slash Y_0$; spectators are suppressed.}:
\begin{enumerate}
    \item $\urm(N^{(\text{CS})})$ connected to overbalanced nodes on the left and right. The non-genuine monopole $\mathfrak{M}_{\mathrm{probe}}$ transforms as a singlet under $\mathfrak{f}_A$ and $\mathfrak{f}_B$: \ie 
    \begin{subequations}
        \label{eq:mix_sym_F_N=4}
    \begin{align}
        p_{\mathrm{CS}}=(0,0)\;.
    \end{align}
    
    \item $\urm(N^{(\text{CS})})$ connected to overbalanced nodes $r\in\Ocal_A$ on the left and to a balanced subset $J\in \Bcal_B$ on the right. The non-genuine monopole $\mathfrak{M}_{\mathrm{probe}}$ transforms as a singlet under $\mathfrak{f}_A$ and as $[1,0,\ldots,0]_{\surm(\ell_J+1)}$ under the $\surmL(\ell_J+1)_B \subset \mathfrak{f}_B$. On a computational level, this requires the map $w^{(\text{CS})} = 1/ b_1$, with $b_1$ the fundamental weight fugacity of $\surmL(\ell_J+1)_B$. Hence, 
    \begin{align}
        p_{\mathrm{CS}}=(0,+1_J) \;.
    \end{align}

    \item Analogously, if $\urm(N^{(\text{CS})})$ is connected to a balanced subset $I\in \Bcal_A$ on the left and to an overbalanced node $r\in\Ocal_B$ on the right, then $\mathfrak{M}_{\mathrm{probe}}$ transforms  as $[0,\ldots,0,1]_{\surm(\ell_I+1)}$ under the $\surmL(\ell_I+1)_A \subset \mathfrak{f}_A$ and trivially under $\mathfrak{f}_B$. Likewise, this requires the map $w^{(\text{CS})} = 1/ a_{\ell_I}$ with $a_{\ell_I}$ the anti-fundamental weight fugacity of $\surmL(\ell_I+1)_A$. Hence, 
    \begin{align}
    p_{\mathrm{CS}}=(-1_I,0)\;.
    \end{align}

    \item $\urm(N^{(\text{CS})})$ connected to a balanced subset $I\in\Bcal_A$ on the left and to a balanced subset $J\in\Bcal_B$ on the right. $\mathfrak{M}_{\mathrm{probe}}$ transforms as $[0,\ldots,0,1]_{\surm(\ell_I+1)} \times [1,0,\ldots,0]_{\surm(\ell_J+1)}$ under the $\surmL(\ell_I+1)_A \oplus \surmL(\ell_J+1)_B \subset \mathfrak{f}_A\oplus  \mathfrak{f}_B$. To achieve this practically, one needs the following map $w^{(\text{CS})} = 1/( a_{\ell_I} b_1)$; see Appendix~\ref{app:CSM_TSUN} for details. Therefore,
      \begin{align}
             p_{\mathrm{CS}}=(-1_I,+1_J)\;.
    \end{align}
    \end{subequations}
\end{enumerate}
Consequently, for any linear 3d $\Ncal=4$ CSM quiver with unitary nodes, one finds:
\begin{itemize}
    \item If $[ p_{\mathrm{CS}}]$ is trivial, then $\Zcal_{\text{tot}}= \Zcal_A \times \Zcal_B$ and $F=\Fcal_A \times \Fcal_B$.
    \item If only one component of $[ p_{\mathrm{CS}}]$ is non-zero, then $\Zcal_{\text{tot}}$ factorises. No $A-B$ quotient is induced, but $F$ may differ from $\Fcal_A \times \Fcal_B$.
    \item If both components are non-trivial, $\Zcal_{\text{tot}}$ is generically a product and can couple the two centres. Let $\alpha =[p_A] \in \widehat{\Zcal}_A$ and $\beta=[p_B] \in \widehat{\Zcal}_B$ have orders $r$ and $s$, respectively. The common coupled subgroup is $\Z_{\gcd(r,s)}$. Thus, a non-trivial diagonal quotient arises only if both classes are non-trivial and $\gcd(r,s)>1$.
\end{itemize}

\subsubsection{2-groups and mixed 't~Hooft anomalies}
\label{sec:mixed_tHooft_N=4_case_general}
The arguments presented so far allow one to determine the global form of the faithful 0-form symmetry group $\Fcal$ and its cover $F$, through the monopole $\mono{\text{probe}}$. This is also a key ingredient to determine a 2-group symmetry between the 1-form symmetry $\Gamma^{(1)}=\Z_\kappa^{(1)}$ and $\Fcal$. Consider a Wilson line $W_{1\otimes \ldots \otimes 1\otimes F \otimes 1 \otimes \ldots \otimes 1}$ in the fundamental representation of the central node of \eqref{eq:example_quiver_symmetries}, that is charged under the 1-form symmetry. It cannot end, but its $\kappa$-fold power does end on a gauge invariant local operator, which is the conjugate of $\mono{\text{probe}}$ used in the previous section. Denote again by $[ p_{\mathrm{CS}}]$ the complete character of this dressed unit monopole.
\begin{itemize}
    \item If $[ p_{\mathrm{CS}}]$ is trivial, the 2-group is trivial, see Examples 1, 2 in Sections~\ref{subsubsec:Ex1} and \ref{subsubsec:Ex2}.
    \item If only one component of $[ p_{\mathrm{CS}}]$ is non-zero, then there is a non-trivial obstruction class for the lifting of $\Fcal$ to $F$. The 2-group is non-trivial with $\Fcal_{A,B}$ depending on which $p_{A,B}$ is non-zero, see Example 3 in Section~\ref{subsubsec:Ex3}.
    \item If both components are non-trivial, the 2-group involves both $\Fcal_{A,B}$ and the subgroup of the centre of $F$ entering the short exact sequence is $\Z_{\lcm rs}$, see Examples 4, 5 in Sections~\ref{subsubsec:Ex4} and \ref{subsubsec:Ex5}.
\end{itemize}

\paragraph{Monopoles with fractional flux}
In order to discuss mixed 't~Hooft anomalies between the 1-form symmetry and the 0-form symmetries, focus on the non-genuine monopole operator $\mono{\text{frac}}$ with fractional flux, introduced in Section \ref{subsubsec:fractional_monopoles}. 
For the theory \eqref{eq:example_quiver_symmetries}, consider the central node $\urm(N^{(\text{CS})})$, which is either connected to an overbalanced node $r$ in the set $\Ocal_X$ or to a balanced component $\Bcal_X$.

In the first case, the associated 0-form symmetry is $\Fcal_X=\urm(1)_X$ and the UV topological fugacity is $w_r^{(X)}$. After the appropriate fugacity maps \eqref{eq:mix_sym_Fcal_N=4}, based on the other components connected to the node $r$, $w_r^{(X)}$ still contains a pure $\urm(1)_X$ fugacity $q_r$. It follows that the charge of $\mono{\text{frac}}$ under $\urm(1)_X$ is given by
\begin{equation}
    \sum_{i=1}^{N_r} \frac{1}{\kappa} = \frac{N_r}{\kappa} \, . 
\end{equation}
The fractional charge is not an allowed charge under $\urm(1)_X$, hence $\mono{\text{frac}}$ transforms in a representation of the $\kappa$-fold cover $\widetilde{\urm(1)}_X$. This is encoded in the mixed 't~Hooft anomaly
\begin{equation}
     \label{eq:mix_tHooft_general}
    S_{\text{mix}} \sim 2 \pi \im \int_{\Mcal} \frac{N_r}{\kappa} B_2 \cup (c^X_1 \, \text{mod} \, \kappa) \, ,
\end{equation}
between the 1-form symmetry and the $\urm(1)_X$ 0-form symmetry.

In the second case, the 0-form symmetry associated to the set of balanced nodes $\Bcal_X$ is a non-abelian $\Fcal_X$. Suppose, $\Fcal_X=\psurm(N)_X$, then a mixed anomaly between $\Fcal_X$ and the 1-form symmetry can be computed only in the case of a split 2-group, namely $\gcd(N,\kappa)=1$. However, the coefficient of an anomaly of the form $B_2 \cup \omega_2(\psurm(N))$ is proportional to $1/\gcd(N,\kappa) = 1$, hence the anomaly is always trivial.
    
Moreover, if between $\Bcal_X$ and the central node there is an overbalanced node $r$, as in Example 5 below, the only surviving mixed anomaly after an appropriate fugacity map is the one with the abelian 0-form symmetry. 
In this case the non-abelian symmetry is not involved in a 2-group with the 1-form symmetry but its mixed anomaly is still trivial.
In fact, the presence of a non-abelian topological symmetry requires constraints on the parameters of the theory, in particular the balance condition. On a set of three adjacent nodes with fugacities $w_j$ the introduction of a uniform shift of the monopole fluxes induces the fugacity factor
\begin{subequations}
    \label{eq:balanced_nodes_no_mix_1-form}
\begin{equation}
    w_{i-1}^{\frac{N_{i-1}}{\kappa}} \, \, w_{i}^{\frac{N_{i}}{\kappa}} \, \, w_{i+1}^{\frac{N_{i+1}}{\kappa}} \, .
\end{equation}
However, the map to the IR fugacities $x_i$ with the Cartan matrix gives: 
\begin{equation}
    x_i ^{\frac{1}{\kappa} \,  \big (2 N_i - N_{i-1} - N_{i+1} \big)} \, , 
\end{equation}
\end{subequations}
in which the exponent is always zero due to the balance condition for the three nodes. One concludes that, in the class of theories considered in this paper, the only possible mixed 't~Hooft anomaly of the 1-form symmetry is with the abelian 0-form symmetry factors.

\subsection{\texorpdfstring{Examples of 3d $\Ncal=4$ CSM theories}{Examples of N=4 CSM theories}}
\label{sec:examples_N=4}
In this section the framework of Section \ref{subsec:general_N=4} is illustrated in five examples of 3d $\Ncal=4$ CSM theories.
The aim of the first two examples is to present theories with $\urm(1)$ 0-form symmetries, hence with trivial 2-group structures. The key feature is the mixed 't~Hooft anomaly between the 0-form factors and the 1-form symmetry.
In the last three examples, one restricts to theories where at least one 0-form factor $\mathcal{F}_A$ or $\mathcal{F}_B$ is non-abelian. The goal is to study the 0-form global forms, the non-trivial 2-group structure and the possible 't~Hooft anomalies.

\subsubsection{Ex.~1: one single abelian 0-form symmetry}
\label{subsubsec:Ex1}
Consider the following brane configuration of two NS5s and one $(1,\kappa)$ 5-brane. 
\begin{equation}
\label{eq:example_1_Unitary}
\begin{minipage}[c]{0.3\textwidth}
    \centering
    \includegraphics[width=0.7\linewidth]{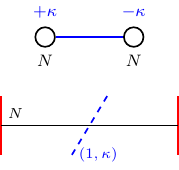}
\end{minipage}
\hspace{-8pt}
\raisebox{1.85em}{$\overset{\text{GK}}{\longleftrightarrow}$}
\hspace{0pt}
\begin{minipage}[c]{0.3\textwidth}
    \centering
    \includegraphics[width=0.85\linewidth]{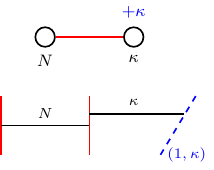}
\end{minipage}
\end{equation}
Focus on the factorised phase, on the right-hand side. 
For $\kappa > 2N$\footnote{To be a \emph{good} CSM theory: $\kappa \geq 2N$. For $\kappa = 2N$, the theory has enhanced non-abelian symmetry $\Fcal_A = \psurm(2)_A$ in the IR, thus a non-trivial 2-group can be realised, see Examples 3, 4 and 5. }, the symmetries are
\begin{align}
    \Fcal_A = \urm(1)_A \,, 
    \quad
    \Fcal_B = \text{trivial }\,, 
    \quad 
    \Gamma^{(1)} = \Z_\kappa^{(1)} \;.
\end{align}
Therefore, the 2-group structure is trivial.
The theory has a mixed 't~Hooft anomaly between $\Fcal_A$ and $\Gamma^{(1)}$:
\begin{align}
    \label{eq:mix_tHooft_Ex1}
    S_{\text{mix}} \sim 2 \pi \im \int_{\Mcal} \frac{N}{\kappa} B_2 \cup (c^A_1 \, \text{mod} \, \kappa) \, ,
\end{align}
where $c^A_1$ is the first Chern class of the $\urm(1)_A$ symmetry. The term $(c^A_1 \, \text{mod} \, \kappa)$ entering the action encodes the obstruction to lift the $\urm(1)_A$ bundle to the $\widetilde{\urm(1)}_A$. 
The anomaly coefficient is understood from the charge of the monopole $\mono{\text{frac}}$ under $\Fcal_A$, as introduced in Section \ref{sec:mixed_tHooft_N=4_case_general}.
Applying the same logic to this example \eqref{eq:example_1_Unitary}, one defines the topological $\urm(1)_A$ fugacity $w_1$ for the $\urm(N)$ gauge node of the factorised phase and finds that $\mono{\text{frac}}$ has fractional $w_1$ charge; this signals the mixed anomaly.  

By construction, the 1-form symmetry is anomaly-free in the theory $T$ given by \eqref{eq:example_1_Unitary}, so it is possible to gauge $\Gamma^{(1)}$ and determine the structure of the 0-form global symmetry group after the gauging, following the logic of \cite{Bergman:2020ifi}. The question is whether the 0-form symmetry $\Z^{(0)}_{\kappa}$, which is the Pontryagin dual of $\Gamma^{(1)}$, is independent of the remaining $\urm(1)_A$ 0-form symmetry. In particular, in the theory $\widetilde{T} = T/\Gamma^{(1)}$ the monopole $\mono{\text{frac}}$ with fractional fluxes becomes a new local operator of the theory with charge $N/\kappa$ under the $\urm(1)_A$ and charge $(1\,\text{mod}\,\kappa)$ under the new $\Z^{(0)}_{\kappa}$. The only independent subgroup of $\Z^{(0)}_{\kappa}$ is then $\Z_{\gcd(N,\kappa)}$, since its action on local operators cannot be realised by the $\urm(1)_A$ symmetry. Then, the 0-form symmetry group is
\begin{equation}
    \Fcal_{\widetilde{T}} = \widetilde{\urm(1)}_A \times \Z_{\gcd(N,\kappa)}
\end{equation}
where the $\widetilde{\urm(1)}_A$ denotes the $\kappa$-fold cover of $\urm(1)_A$ if $g =\gcd(N,\kappa)=1$ and the $\kappa'$-fold cover of $\urm(1)_A$ if $\gcd(N,\kappa)>1$, with $\kappa= \kappa' \cdot g$. In fact, the mixed 't~Hooft anomaly \eqref{eq:mix_tHooft_Ex1} implies that after the gauging of the 1-form symmetry, the 0-form symmetry group global form is fixed to be the one that trivialises the obstruction $(c_1^A\,\text{mod}\,\kappa)$, namely $\widetilde{\urm(1)}_A$.

\subsubsection{Ex.~2: two abelian 0-form symmetries}
\label{subsubsec:Ex2}
Consider the following theory given by two NS5s and two $(1,\kappa)$ 5-branes.
\begin{equation}
\label{eq:example_2_Unitary}
\begin{minipage}[c]{0.4\textwidth}
    \centering
    \includegraphics{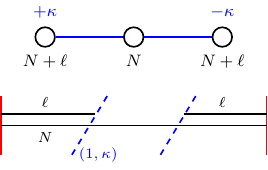}
\end{minipage}
\hspace{5pt}
\raisebox{2em}{$\overset{\text{GK}}{\longleftrightarrow}$}
\hspace{10pt}
\begin{minipage}[c]{0.4\textwidth}
    \centering
    \includegraphics{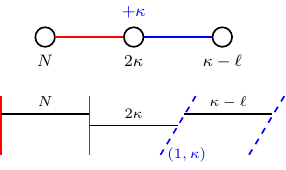}
\end{minipage}
\end{equation}
It is parametrised by an integer $\ell < \kappa$.
For $\kappa > N +\ell$ and $\ell>0$, the symmetries are 
\begin{align}
    \Fcal_A = \urm(1)_A \;, \quad
    \Fcal_B= \urm(1)_B \;, \quad
    \Gamma^{(1)} = \Z_\kappa^{(1)}
\end{align}
and there is again only a trivial 2-group. The factorised phase of the theory offers a clean way to read off the $\Fcal_A$ and $\Fcal_B$ symmetries: they are the topological symmetry factors associated, respectively, to the $\urm(N)$ and $\urm(\kappa - \ell)$ gauge nodes.
The 4d anomaly action is
\begin{align}
    S_{\text{mix}} \sim 2 \pi \im \int_{\Mcal} 
    \frac{N}{\kappa} B_2 \cup (c_1^A \, \text{mod} \, \kappa)
    +\frac{N}{\kappa} B_2 \cup (c_1^B \, \text{mod} \, \kappa)
\end{align}
for mixed 't~Hooft anomalies between the 1-form symmetry, with background $B_2$, and the $A$ and $B$-type 0-form symmetries, with Chern classes of background field strengths $c_1^A$ and $c^B_1$, respectively. 

Note that there is no pure 0-form mixed anomaly between $\urm(1)_A$ and $\urm(1)_B$. A Wilson line inducing the background for $\urm(1)_A$ can end on a monopole with non-trivial flux under the central node of the theory on the right of \eqref{eq:example_2_Unitary}.
Consider the UV fugacities $w_j$ associated to every gauge node; to get the IR fugacity of the central node, no mixing between the $w_j$ is required in the abelian case. Hence, the monopole under consideration carries no charge under $\urm(1)_B$.

As in the previous example, one can gauge the 1-form symmetry, which leads to a $\urm(1)_A \times \urm(1)_B \times \Z_{\gcd(N,\kappa)}$ 0-form symmetry in the theory after gauging.

\subsubsection{Ex.~3: one abelian and one non-abelian 0-form symmetry}
\label{subsubsec:Ex3}
Consider the following CSM theory given by two NS5s and $M$ $(1,\kappa)$ 5-branes.
\begin{equation}
\label{eq:example_3_Unitary}
\begin{minipage}[c]{0.45\textwidth}
    \centering
    \includegraphics[width=\linewidth]{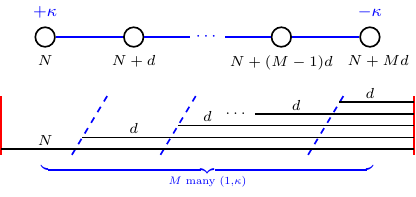}
\end{minipage}
\hspace{5pt}
\raisebox{2.5em}{$\overset{\text{GK}}{\longleftrightarrow}$}
\hspace{5pt}
\begin{minipage}[c]{0.475\textwidth}
    \centering
    \raisebox{0.15\height}{\includegraphics[width=1\linewidth]{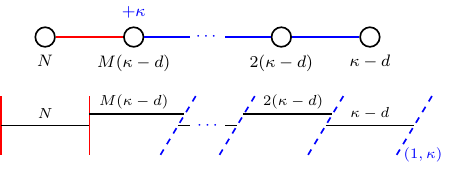}}
\end{minipage}
\end{equation}
It is parametrised by $0 \leq d\leq\kappa$. The symmetries of the theory are
\begin{align}
\label{eq:symm_Ex3}
    \Fcal_A =\urm(1)_A \;,\quad 
    \Fcal_B = \psurm(M)_B \;,\quad 
    \Gamma^{(1)} = \Z_\kappa^{(1)}\; .
\end{align}
Note the enhancement of $\Fcal_B$ to a non-abelian symmetry in the IR due to the sequence of balanced nodes.

\paragraph{Case $d = 0$: non-trivial 2-group and mixed anomaly.}
To study the non-trivial 2-group structure between $\Fcal_B$ and $\Gamma^{(1)}$, consider the monopole operator sitting at the end of $(W_{1 \otimes \ldots \otimes F \otimes \ldots \otimes 1})^{\otimes \kappa}$ in the factorised phase of \eqref{eq:example_3_Unitary}. As explained in Section \ref{sec:mixed_tHooft_N=4_case_general}, this monopole, together with others, sits in the fundamental representation $[1,0,\ldots,0]_{A_{M-1}}$ of $\surm(M)_B$, which is not an allowed representation of $\Fcal_B$. 
The connected group 
\begin{equation}
    F=  \urm(1)_A \times \surm(M)_B 
\end{equation}
is the cover of the faithful 0-form symmetry group 
\begin{align}
    \Fcal = \frac{F}{\Zcal} = \urm(1)_A \times \frac{\surm(M)_B}{\Z_M} \qquad \text{with} \qquad \mathcal{Z}= \Z_M \subset Z(F) = \urm(1)_A \times \Z_M \, ,
\end{align}
that takes part in the short exact sequence $ 0\to \Z_\kappa \to \mathcal{E} \to \mathcal{Z} \to 0$. The abelian factor does not enter the 2-group, because the monopole operator considered is not charged under $\urm(1)_A$; in other words, $\Zcal = \Z_M$ in the quotient only acts on $\surm(M)_B$. 

Suppose $\gcd(\kappa,M)>1$, then the 2-group is non-trivial and $\Ecal=\Z_{\kappa M}$. Consequently, the mixed 't~Hooft anomaly involves the 2-group and $\Fcal_A$:
\begin{align}
\label{ex3anom2group}
     S^{\text{2-group}}_{\text{mix}} \sim 2 \pi \im \int_{\Mcal} 
    \frac{N}{\kappa M} \widetilde{B}_2 \cup (c_1^A \, \text{mod} \, \kappa M) \,,
\end{align}
with the $\Z_{\kappa M}$-valued 2-group background $\widetilde{B}_2$. 

Suppose, instead, $\gcd(\kappa,M)=1$, namely a trivial 2-group; define the 2-group background as $\widetilde{B}_2 = M B_2 + \widetilde{\omega}_2^B$, by taking the lifts to $\Z_{\kappa M}$ of the 1-form symmetry background $B_2$ valued in $\Z_{\kappa}$ and of the obstruction class $\omega_2^B = \omega_2(\psurm(M))$ valued in $\Z_M$. The 't~Hooft anomaly (\ref{ex3anom2group}) reduces to the mixed anomaly between the 1-form symmetry and $\mathcal{F}_A$ and the mixed anomaly between $\mathcal{F}_B$ and $\mathcal{F}_A$:
\begin{equation}
\label{eq:N=4_CSM_Ex3_mix_anomaly}
\begin{aligned}
    S_{\text{mix}} & \sim 2 \pi \im \int_{\Mcal} 
    \frac{N M}{\kappa M} \,  B_2 \cup (c_1^A  \, \text{mod} \, \kappa M ) + 2 \pi \im \int_{\Mcal} \frac{N}{\kappa M} \, \widetilde{\omega}_2^B \cup (c_1^A \, \text{mod}\, \kappa M) \\
    & \sim 2 \pi \im \int_{\Mcal} 
    \frac{N}{\kappa } \,  B_2 \cup (c_1^A  \, \text{mod} \, \kappa ) 
    + 2 \pi \im \int_{\Mcal} \frac{N}{M} \, \omega_2^B \cup (c_1^A \, \text{mod}\, M) \, .
\end{aligned}
\end{equation}
Going from the first to the second line shows how to reduce the lift $\widetilde{\omega}_2^B$ to the $\Z_M$ valued class $\omega_2^B$.
It follows that gauging the 1-form symmetry produces, as highlighted in the first two examples, a theory with the 0-form symmetry group: $\urm(1)_A \times \Z_{\gcd(N,\kappa)} \times \psurm(M)_B$. 

\paragraph{Case $d\neq 0$: non-trivial 2-group and self-anomaly.}
In the absence of the background of the 0-form symmetries, it is possible to study the non-trivial self-anomaly for the 1-form symmetry $\Z_{\kappa}^{(1)}$. As discussed in Section~\ref{subsubsec:fractional_monopoles} the anomaly is given by
\begin{equation}
    S^{\text{1-form}}_{\text{self}} \sim 2\pi\im  \int_{\mathcal{M}_4} \frac{M d}{\kappa} B_2 \cup B_2
    \, .
\end{equation}
Hence, the anomaly-free subgroup of the 1-form symmetry is $\Z_{\gcd(Md,\kappa)}^{(1)}$. Its gauging introduces in the new theory the dual 0-form symmetry $\Z_{\gcd(Md,\kappa)}^{(0)}$.

The 2-group structure remains the same as in the case $d=0$, together with the anomaly in \eqref{ex3anom2group}. However, an additional piece of data is the self-anomaly of the non-trivial 2-group symmetry given by
\begin{equation}
\label{eq:ex3_2-group_self}
    S_{\text{self}}^{2-\text{group}} \sim 2 \pi \im \int_{\Mcal} 
    \frac{d}{\kappa M} \, \widetilde{B}_2 \cup \widetilde{B}_2 \, .
\end{equation}
The anomaly action explicitly depends on $d$ because the 2-group line operator carries precisely charge $(d \mod \kappa M)$ under $\Ecal$.
The parameter $d$ cleanly distinguishes between theories in which both the 2-group self-anomaly and the 1-form self-anomaly are trivial ($d=0$) or non-trivial ($d\neq 0$). 

Additionally, when the 2-group splits, define $\widetilde{B}_2 = M B_2 + \, \widetilde{\omega}_2^B$ to see that the self-anomaly \eqref{eq:ex3_2-group_self} reduces to the 1-form self-anomaly, the trivial mixed 1-form and 0-form anomaly and the 0-form self-anomaly:
\begin{equation}
    S_{\text{split}} \sim 2 \pi \im \int_{\Mcal} 
    \frac{M d}{\kappa} \, B_2 \cup B_2 + 2 d B_2 \cup 2 \omega_2^B + \frac{d}{M} \omega_2^B \cup \omega_2^B \, .
\end{equation}
Note that it is always possible to shift the total 2-group self-anomaly to rewrite the coefficient as $M(\kappa -d)$, matching the 1-form self-anomaly coefficient of the GK-dual framework. The advantage is that the $\psurm(M)_B$ self-anomaly then is given by the integral of $\frac{\kappa -d}{M} \omega_2^B \cup \omega_2^B $ which agrees with the computation of the charge under $\Z_M$ of a line inducing the background of the $\surm(M)_B$ symmetry.

\subsubsection{Ex.~4: two equal non-abelian 0-form symmetries}
\label{subsubsec:Ex4}
Consider the following CSM theory, with $N$ NS5 branes and $N$ $(1,\kappa)$ 5-branes.
\begin{gather}
    \raisebox{-.5\height}{\includegraphics[width=0.8\linewidth]{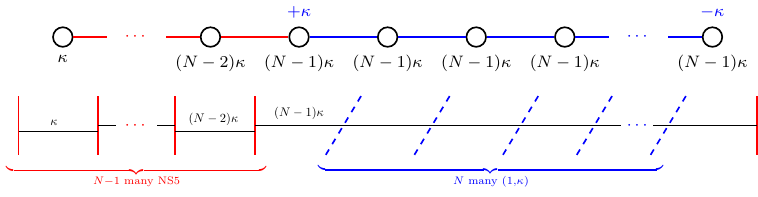}} 
    \nonumber\\
    \label{eq:example_4_Unitary}
    \qquad\Big\updownarrow\text{GK}
    \\
    \raisebox{-.5\height}{\includegraphics[width=0.8\linewidth]{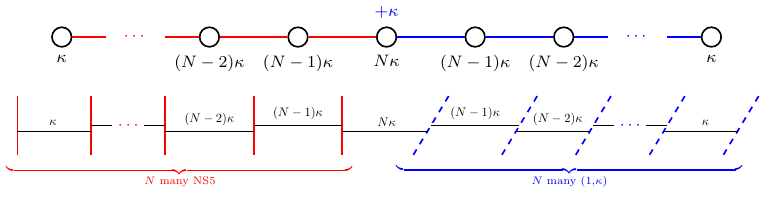}}
    \nonumber
\end{gather}
The symmetry data is 
\begin{align}
\Fcal_A =\psurm(N)_A \,, \quad
\Fcal_B = \psurm(N)_B \,, \quad
\Gamma^{(1)}= \Z_\kappa^{(1)} \,.
\end{align}
The dressed monopole operators on which $(W_{1\otimes \ldots \otimes 1\otimes F \otimes 1 \otimes \ldots \otimes 1})^{\otimes \kappa}$ can end transform in the representation $[0,\ldots,0,1]_A \otimes [1,0,\ldots,0]_B$ of $\surm(N)_A \times \surm(N)_B$, see also Appendix~\ref{app:CSM_TSUN}. They are non-genuine local operators, that are invariant under the $\Z^{\text{diag}}_N$ obtained as a diagonal combination of the centres of the two $\surm(N)$.
This implies that the global form of the symmetry acting both on genuine and non-genuine local operators is
\begin{equation}
\label{eq:quotient_cover}
    F=\frac{\surm(N)_A\times \surm(N)_B}{\Z^{\text{diag}}_N} \, .
\end{equation}
$\Fcal = F/ \mathcal{Z}$ defines the group $\mathcal{Z} = \Z^{\text{anti-diag}}_N$, which is the one entering the short exact sequence $0 \to \Gamma^{(1)} \to \mathcal{E} \to \mathcal{Z} \to 0$, that characterises the 2-group between $\Gamma^{(1)}$ and the entire $\Fcal = \Fcal_A \times \Fcal_B$.

When the 2-group is non-trivial, namely $\gcd(\kappa,N)>1$, the only possible 't~Hooft anomaly is a self-anomaly of the 2-group and there are no mixed anomalies because all the 0-form symmetries are involved in the 2-group. The coefficient of the self-anomaly is given by the charge $q_{\mathcal{E}}$ under $\mathcal{E}$ of the monopole on which $(W_{1\otimes \ldots \otimes 1\otimes F \otimes 1 \otimes \ldots \otimes 1})^{\otimes \kappa}$ can end. The charge $q_{\mathcal{E}}$ is obtained combining the charge under the 1-form symmetry, namely the gauge charge $q$ mod $\kappa$, and the charge $q_{\Zcal}$ under $\Zcal=\Z_N$. The result is
\begin{equation}
    q_{\mathcal{E}} = \kappa + \kappa (N-1) = \kappa N = 0 \bmod \kappa N \, ,
\end{equation}
which implies a trivial 2-group self-anomaly. Turning off the background of the 0-form symmetries, one notices that also the 1-form symmetry is anomaly-free, since the ranks of the gauge nodes are always proportional to the CS-level $\kappa$. 

After gauging the anomaly-free 1-form symmetry, the resulting theory is endowed with the 0-form symmetry group $\psurm(N)_A \times \psurm(N)_B \times \Z_{\kappa}$. Since the mixed anomaly between the 1-form and the 0-form symmetries is trivial, there is no non-trivial extension of the 0-form global symmetry in the gauged theory.

\paragraph{}  As a remark, for $\kappa=1$ the CSM theory is dual to the standard 3d $\Ncal=4$ $T[\surm(N)]$ theory without any 1-form symmetry. This theory presents a mixed anomaly between the two 0-form symmetries $\Fcal_A$ and $\Fcal_B$ given by
\begin{equation}
       S_{\text{mixed}} \sim 2 \pi \im \int_{\Mcal} \frac{1}{N} \, \omega_2^A \cup \omega_2^B \, , 
\end{equation}
where the classes $\omega_2^{A,B}$ obstruct the lift to $\surm(N)_{A,B}$ respectively. With a different choice of the cover group $F$ this result is recovered in the CSM theory \eqref{eq:example_4_Unitary}. The global form \eqref{eq:quotient_cover} is the minimal choice, for which all the representations are realised by genuine and non-genuine local operators. It defines explicitly the $\Zcal$ that enters in the definition of the 2-group. Another choice for $F$ is $F=\surm(N)_A \times \surm(N)_B$, from which one reads the definition of the obstruction classes $\omega_2^{A,B}$ and determines the mixed 0-form anomaly. The coefficient is given by the charge under $\Z_N^B$ of the endpoint of a Wilson line in determinant representation inducing the background for $\surm(N)_A$.

\subsubsection{Ex.~5: two different non-abelian 0-form symmetries} 
\label{subsubsec:Ex5}
A generalisation of Example 4 is given by a brane system with $L$ many NS5 and $R$ many $(1,\kappa)$ branes arranged as follows. 
\begin{gather}
    \raisebox{-.5\height}{\includegraphics[width=0.8\linewidth]{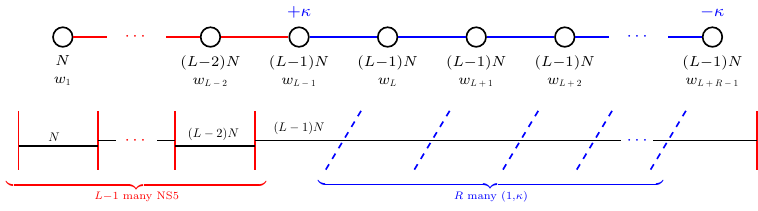}} 
    \nonumber\\
    \label{eq:example_5_Unitary}
    \qquad\Big\updownarrow\text{GK}
    \\
    \raisebox{-.5\height}{\includegraphics[width=0.8\linewidth]{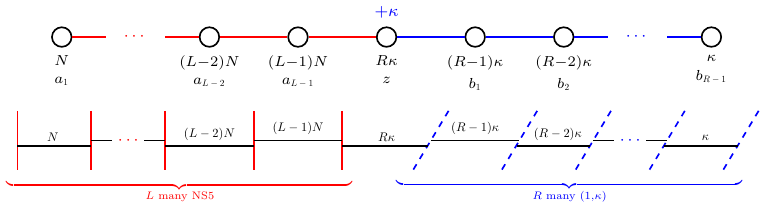}}
    \nonumber
\end{gather}
Note that the number of NS5 and $(1, \kappa)$ branes is different and on the left side the ranks of the gauge nodes are no longer proportional to the CS-level $\kappa$. To be a \emph{good} CSM theory, the condition $LN \leq R\kappa$ is required.
With the IR enhancement due to the balance node condition, the Lie algebra of 0-form global symmetries is
\begin{align}
    \mathfrak{f}_A = \mathfrak{su}(L-1)_A \oplus \mathfrak{u}(1)_A \,, 
    \qquad 
    \mathfrak{f}_B = \mathfrak{su}(R)_B \,. 
\end{align}

The global form of the Lie groups depends on the choice of $N$, $L$, $R$ and $\kappa$. In order to understand this, it is possible to start from the UV mixed anomalies between the 1-form symmetry and the product of $(L-1+R)$ topological $\urm(1)$ symmetry factors. Assigning $\urm(1)$ fugacities $w_j$ to every gauge node in the theory on top of \eqref{eq:example_5_Unitary}, the anomaly action is given by: 
\begin{equation}
    S_{\text{mix}}^{\mathrm{UV}} \sim 2\pi\im \int_{\Mcal} \frac{1}{\kappa} B_2 \cup \Bigg( \sum_{l=1}^{L-2} l \, N \,  c_1(w_l) \, \text{mod}\, \kappa + \sum_{j=L-1}^{L-1+R} (L-1)N \, c_1(w_j) \, \text{mod}\, \kappa  \Bigg) \, .
\end{equation}
The non-trivial fugacity map to the factorised phase fugacities reads 
\begin{equation}
    w_i = a_i   \, , \quad w_{L-1} = a_{L-1} z \, , \quad w_{L-1+j} = b_j  \, , \quad w_{L-1+R} = \frac{1}{ z \,  \Pi_j b_j }
\end{equation}
with $i \in \{1, \ldots, L-2\}$ and $j \in \{1, \ldots, R-1\}$. $z$ is not a symmetry fugacity of the physical spectrum, since only non-genuine operators are charged under the topological $\urm(1)$ of the CS node.
The mixed anomaly in the new fugacities reduces to 
\begin{equation}
\label{eq:example_5_mix}
    S_{\text{mix}}^{\mathrm{UV}} \sim 2\pi\im \int_{\Mcal} \frac{N}{\kappa} B_2 \cup \Bigg( \sum_{l=1}^{L-2} l \, c_1(a_l) \, \text{mod}\, \kappa + (L-1) \,c_1(a_{L-1}) \, \text{mod}\, \kappa  \Bigg) \, .
\end{equation}
The sum of the Chern classes realises the class of the $\urm(L-1)_A$ bundle, which is the global form of the 0-form symmetry in the theory for $LN < R\kappa$.
However, the anomaly can be additionally simplified taking into account the proper map between UV and IR symmetries. In fact, the $\urm(1)_A$ fugacity $a_{L-1}$ needs to mix with the non-abelian $\surm(L-1)_A$ fugacities of the set of adjacent balanced nodes, as in \cite{Nawata:2023rdx,Marino:2025uub}: 
\begin{equation}
a_{L-1} = a \, \bigg(\prod_{l=1}^{L-2} a_l^{\frac{l}{L-1}} \bigg)^{-1} \, ,
\end{equation}
since the monopole charged under the $(L-1)$-th gauge node is part of the fundamental representation of $\surm(L-1)_A$.
The only mixed anomaly that survives is the one between the 1-form symmetry $\Z_{\kappa}$ and the 0-form symmetry $\urm(1)_A$:
\begin{equation}
    S_{\text{mix}}^{\mathrm{IR}} \sim 2\pi\im \int_{\Mcal} \frac{(L-1)N}{\kappa} B_2 \cup  c_1(a) \, \text{mod}\, \kappa \, .
\end{equation}

\paragraph{Symmetry enhancement.} An interesting feature of this example is the possibility of the enhancement of the symmetry
\begin{equation}
    \frac{\surm(L-1)_A \times \urm(1)_A}{\Z_{L-1}} \cong \urm(L-1)_A \; \longrightarrow \; \surm(L)_A \,.
\end{equation}
In order to make the $(L-1)$-th node balanced, it is necessary to introduce a condition on the ranks of the gauge nodes, namely: $LN=R\kappa$. Then, starting from the mixed anomaly \eqref{eq:example_5_mix}, the sum of the Chern classes can be recognised as the second Stiefel--Whitney class for the $\psurm(L)$ bundle
\begin{equation}
    \sum_{l=1}^{L-1} l \, c_1(a_l) \mod L = \omega_2(\psurm(L)) \, .
\end{equation}
The mixed anomaly follows as
\begin{equation}
    S_{\text{mix}}^{\mathrm{IR}} \sim 2\pi\im \int_{\Mcal} \frac{N}{\kappa} B_2 \cup  L \, \omega_2(\psurm(L)) \,\, \text{mod}\, \kappa 
\end{equation}
and, given the fact that $L N$ is proportional to $\kappa$, the anomaly is trivial. This is precisely the same feature of Example 4, in which the ranks are chosen proportional to the CS-level by construction. 

\paragraph{2-group symmetry.}
Considering now the theory in \eqref{eq:example_5_Unitary} with the condition $LN=R\kappa$, the symmetries are:
\begin{align}
\Fcal_A =\psurm(L)_A \,, \quad
\Fcal_B = \psurm(R)_B \,, \quad
\Gamma^{(1)}= \Z_\kappa^{(1)} \,.
\end{align}
One observes that $(W_{1\otimes \ldots \otimes 1\otimes F \otimes 1 \otimes \ldots \otimes 1})^{\otimes \kappa}$ can end on a monopole operator with flux $(\circ,-1)$ under the central node in the factorised phase. It transforms as $[0,\ldots,0,1]_{A_{L-1}}\otimes [1,0,\ldots,0]_{A_{R-1}}$, i.e.\ bifundamentally under $\surm(L)_A \times \surm(R)_B$. The centre $\Z_L\times \Z_R$ acts on the operator as $(\zeta_L^{-1},\zeta_R)$. Denote $\ell =\lcm{L}{R}$, then $ (\zeta_L^{-1},\zeta_R)^\ell = (1,1)$. Therefore, the expected 2-group structure arises from
\begin{equation}
\label{eq:shortexact_Ex5}
    0\to \Z^{(1)}_\kappa \to \Ecal \to \Z_\ell \to 0 \, ,
\end{equation}
in which $\Z_\ell \subset \Z_L\times \Z_R$ is the anti-diagonal combination of the two centres.
The choice for the cover of the faithful symmetry $\Fcal=\Fcal_A \times \Fcal_B$ is
\begin{align}
    F=\frac{\surm(L)_A\times \surm(R)_B}{\Z_{\gcd(L,R)}} \, .
\end{align}
All representations of $F$ are realised by genuine and non-genuine local operators of the theory, since they are all invariant under $\Z_{\gcd(L,R)}$ which is the diagonal combination of $\Z_L$ and $\Z_R$. In this way, it is clear how to obtain from $\Fcal$ and $F$ the definition $\mathcal{Z} = \Z_\ell$ which enters \eqref{eq:shortexact_Ex5} and under which the non-genuine local operators are charged. The 2-group is non-trivial when $\gcd(\ell, \kappa) > 1$ and $\mathcal{E} = \Z_{\ell \kappa}$. The characteristic class $\omega_2$, which is $\Z_\ell$-valued, encodes the obstruction to lift $\Fcal$ to $F$. As in Example 4, the only possible 't~Hooft anomaly, the self-anomaly of the 2-group, is trivial.

\subsection{\texorpdfstring{0-form and 2-group symmetries in $\Ncal=3$ CSM}{0-form and 2-group symmetries in N=3 CSM}}
\label{subsec:sym_structure_N=3}
In this section, the symmetry structures of $\Ncal=3$ CSM quivers are analysed. A key difference is that 0-form symmetries no longer split into A or B type, since the R-symmetry is a single $\surm(2)_R$. Instead the 0-form symmetry contains factors from all maximal branches. 

\subsubsection{Global form of the 0-form symmetry group}
\label{subsubsec:0-form_sym_N=3}
Consider a $\Ncal=3$ CSM linear quiver in the factorised phase of the brane configuration, involving $N_i$ many $(1,\kappa_i)$ 5-branes (ordered from left to right) for $i\in \{1,2,\ldots,\ell\}$. Without loss of generality, one can arrange the brane systems in a factorised phase that satisfies $\kappa_1< \kappa_2 <\ldots <\kappa_\ell$. 
The CS-levels are $k_i=\kappa_{i+1}-\kappa_i > 0 $.
\begin{align}
    \includegraphics[width=\linewidth]{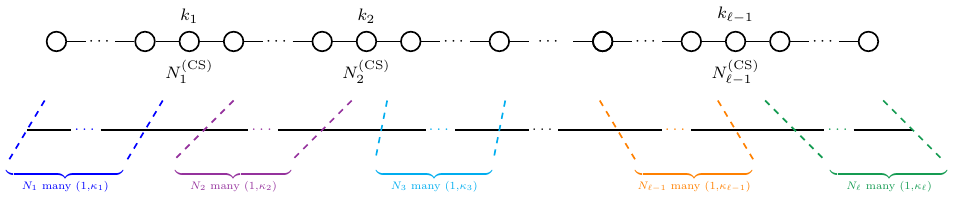}
    \label{eq:gen_N=3_sym_structure_ex}
\end{align}
Based on the brane system and the insights gained in Section~\ref{subsec:general_N=4}, the 0-form symmetry algebra is composed as follows:
each stack of $N_i$ $(1,\kappa_i)$ 5-branes gives rise to the maximal branch isometry algebra $\ffrak_{(1,\kappa_i)}$. 
Explicitly, $\ffrak_{(1,\kappa_i)}$ may contain non-abelian $\surmL$ factors as well as abelian factors\footnote{In the CSM quiver of the factorised phase, the $N_i-1$ many gauge nodes in between the $(1,\kappa_i)$ 5-branes determine abelian vs non-abelian symmetries via balance condition.}, but they are determined analogously to the $f_X$ of \eqref{eq:def_ffrak_X}.

In addition, as a new feature of $\Ncal=3$ theories, there exist $\ell-2$ many additional $\urmL(1)_{\Ocal_i}$ symmetries. Their charged operators $\Ocal_i$ are constructed schematically as follows. Firstly, define $g_{i}=\gcd(k_i,k_{i+1})$ and consider the monopole operator
\begin{align}
\label{eq:N=3_ex_monopole}
\Ocal_i: \qquad 
  \mono{(\bullet,\ldots,\bullet,(k_{i+1}/g_i,\circ),\bullet,\ldots,\bullet,(\circ,-k_{i}/g_i),\bullet,\ldots,\bullet)}  
\end{align}
having trivial magnetic flux at every non-CS node, flux $(k_{i+1}/g_i,\circ)$ at the $\urm(N_i^{(\CS)})_{k_i}$ node and flux $(\circ,-k_{i}/g_i)$ at the $\urm(N_{i+1}^{(\CS)})_{k_{i+1}}$ node. Hence, this is a gauge-invariant bare monopole operator that can be dressed by the matter fields to define $\Ocal_i$, see also Appendix~\ref{app:dressed_monopoles_N3}. This suggests that the $\urm(1)_{\Ocal_i}$ factor has fugacity 
\begin{subequations}
\label{eq:extra_U1}
    \begin{align}
 y_i= (w_{i}^{(\CS)})^{k_{i+1}/g_i} (w_{i+1}^{(\CS)})^{-k_i/g_i} \,,
\end{align}
and its charge $q_i$ is computed on the physical spectrum\footnote{\ie explicitly using the Gauß law constraint:  $k_i\sum_{j=1}^{N_i^{(\CS)}}  m_j^{(i,\CS)} + k_{i+1} \sum_{j=1}^{N_{i+1}^{(\CS)}} m_j^{(i+1,\CS)}=0$.} via
\begin{align}
\label{eq:extra_U1_charge}
    q_i = \frac{g_i}{2} \left( \frac{ \sum_{j=1}^{N_i^{(\CS)}} m_j^{(i,\CS)} }{ k_{i+1} } - \frac{ \sum_{j=1}^{N_{i+1}^{(\CS)}} m_j^{(i+1,\CS)} 
    }{
    k_i
    } \right) \,.
\end{align}
\end{subequations}
Iterating this procedure allows one to define $\ell-2$ many extra $\urm(1)_{\Ocal_i}$ due to adjacent CS-nodes.

In summary, the 0-form symmetry has Lie algebra 
\begin{align}
    \ffrak= \bigoplus_{i=1}^\ell \ffrak_{(1,\kappa_i)} \; \oplus \; \bigoplus_{j=1}^{\ell-2}\urmL(1)_{\Ocal_j}
\end{align}
with covering group $F_0$ and centre $Z_0$
\begin{align}
    F_0 = \prod_{i=1}^\ell F_{(1,\kappa_i)} \; \times \; \prod_{j=1}^{\ell-2}\urm(1)_{\Ocal_j}  \; ,\qquad
    Z_0= \prod_{i=1}^\ell Z_{(1,\kappa_i)} \; \times \; \urm(1)^{\ell-2}
\end{align}
wherein $F_{(1,\kappa_i)}$ denotes the universal cover of $\ffrak_{(1,\kappa_i)}$ and $Z_{(1,\kappa_i)} = Z(F_{(1,\kappa_i)})$ the centre. 

\paragraph{Faithful 0-form symmetry.}
Based on the lessons learnt from Section~\ref{subsec:general_N=4}, the monopole operator $\Ocal_i$~\eqref{eq:N=3_ex_monopole} transforms as follows:
\begin{itemize}
    \item 
    $[0,\ldots,0,k_{i+1}/g_i]_{(1,\kappa_i)}$ under some non-Abelian $\surmL$ of the $(1,\kappa_i)$ branch if
    the node to the left of $\urm(N_i^{(\CS)})_{k_i}$ is balanced, otherwise as charge $k_{i+1}/g_i$ under an abelian factor of this branch isometry.
    \item
    $[k_{i+1}/g_i,0,\ldots,0]_{(1,\kappa_{i+1})}$ under some non-Abelian $\surmL$ of the $(1,\kappa_{i+1})$ branch if the node to the right of $\urm(N_i^{(\CS)})_{k_i}$ is balanced, otherwise as charge $k_{i+1}/g_i$ under an abelian factor of the same branch isometry. 
    \item
    $[k_{i}/g_i,0,\ldots,0]_{(1,\kappa_{i+1})}$ under some non-Abelian $\surmL$ of the $(1,\kappa_{i+1})$ branch if the node to the left of $\urm(N_{i+1}^{(\CS)})_{k_{i+1}}$ is balanced, otherwise as charge $-k_{i}/g_i$ under an abelian factor of the same branch isometry.
    \item 
    $[0,\ldots,0,k_{i}/g_i]_{(1,\kappa_{i+2})}$ under some non-Abelian $\surmL$ of the $(1,\kappa_{i+2})$ branch if the node to the right of $\urm(N_{i+1}^{(\CS)})_{k_{i+1}}$ is balanced, otherwise as charge $-k_{i}/g_i$ under an abelian factor of the same branch isometry.
\end{itemize}
Based on this data, the centre charges $q_{\Ocal_i}$ of all $\Ocal_i$ can be derived, and the faithful 0-form symmetry is obtained by defining
\begin{align}
    Y_0 = \bigoplus_{r=1}^\ell Y_{(1,\kappa_r)} \oplus \mathrm{Span}_\Z\{ q_{\Ocal_i} | \; i=1,\ldots,\ell-2 \} 
\end{align}
where the $Y_{(1,\kappa_r)}$ are the analogous objects as $Y_X$ of \eqref{eq:def_Y_X} for the $(1,\kappa_i)$ branches.
Then one finds $\widehat{\Zcal}_0 = \widehat{Z}_0 \slash Y_0$ with Pontryagin dual $\Zcal_0 =(\widehat{\Zcal}_0)^\vee$ such that $\Fcal = F_0 \slash \Zcal_0$.

\paragraph{Non-genuine operators and extended group.}
For the configuration~\eqref{eq:gen_N=3_sym_structure_ex}, there are $\ell-1$ guiding types of non-genuine monopole operators, \ie the unit monopoles at the $\ell-1$ different CS-nodes:
   \begin{align}
       \label{eq:N=3_ex_mono-probes}
    \mono{\mathrm{probe},i} \equiv \mono{(\bullet,\ldots,\bullet,(1,\circ),\bullet,\ldots,\bullet)} \,, 
    \end{align} 
    which has flux $(1,\circ)$ at $\urm(N_i^{(\CS)})_{k_i}$ and trivial magnetic flux at all other gauge nodes. Their symmetry charges are straightforwardly derived analogously to the above:
\begin{itemize}
    \item 
    $\mono{\mathrm{probe},i}$ transforms as $[0,\ldots,0,1]_{(1,\kappa_i)}$ under some non-Abelian $\surmL$ of the $(1,\kappa_i)$ branch if the node to the left of $\urm(N_i^{(\CS)})_{k_i}$ is balanced, otherwise as charge $1$ under an abelian factor of the same branch isometry.
    \item 
    $\mono{\mathrm{probe},i}$ transforms as $[1,0,\ldots,0]_{(1,\kappa_{i+1})}$ under some non-Abelian $\surmL$ of the $(1,\kappa_{i+1})$ branch if the node to the right of $\urm(N_i^{(\CS)})_{k_i}$ is balanced, otherwise as charge $1$ under an abelian factor of the same branch isometry. 
\end{itemize}
This defines the centre charges $q_{\mathrm{probe},i}$ of $\mono{\mathrm{probe},i}$ and the extended 0-form group is computed by extending
\begin{align}
    Y_{\mathrm{tot}} =  Y_0  \oplus \mathrm{Span}_\Z\{ q_{\mathrm{probe},i} | \; i=1,\ldots,\ell-1 \} 
\end{align}
and likewise adjusting the quotient $\widehat{\Zcal}_{\mathrm{tot}} = \widehat{Z}_0 \slash Y_{\mathrm{tot}} $ with Pontryagin dual $\Zcal_{\mathrm{tot}} = (\widehat{\Zcal}_{\mathrm{tot}})^\vee$ such that $F = F_0 \slash \Zcal_{\mathrm{tot}}$.

\subsubsection{2-groups and mixed 't~Hooft anomalies} 
The existence of a non-trivial 2-group depends directly on the centre charges for the probes $\mono{\mathrm{probe},i}$. If the $i$-th probe transforms non-trivially under the centre of the maximal branch isometries $\Fcal_{(1,\kappa_i)}$ or $\Fcal_{(1,\kappa_{i+1})}$ then the non-abelian factors therein generically participate in the 2-group. Importantly, abelian factors do not give rise to a non-trivial 2-group. Hence, none of the extra $\urm(1)_{\Ocal_i}$ contribute.
A consequence is that a linear $\Ncal=3$ CSM can support a non-trivial 2-group that can involve all maximal branch isometries, provided they support suitable non-abelian factors. 

In the case of a trivial or split 2-group, the mixed 1-form 0-form 't~Hooft anomalies are readily deduced:
\begin{itemize}
 \item For branch isometries $\Fcal_{(1,\kappa_i)}$, the mixed anomalies between non-abelian 0-form factors and the 1-form symmetry are trivial, by the same arguments as in Section~\ref{sec:mixed_tHooft_N=4_case_general}.

 Only abelian branch isometries can have mixed anomalies with $\Gamma^{(1)} = \Z_h^{(1)}$, which are generically of the type $\int \frac{\#}{h} B_2 \cup c_1$ with $\#$ set by the rank of the gauge nodes supporting the abelian isometry.

 \item The mixed anomaly between $\Z_{h}^{(1)}$ and $\urm(1)_{\Ocal_i}$ is read off from the charge formula~\eqref{eq:extra_U1_charge}, applied to gauge-invariant monopole operators with  $\frac{1}{h}$-fractionally shifted coweights. Since the Gauß law constraint holds also for those operators, their $\urm(1)_{\Ocal_i}$ charge is always integer valued. Thus, no mixed anomaly arises.
\end{itemize}

\subsection{\texorpdfstring{Illustration for 3d $\Ncal=3$ CSM quivers}{Illustration for 3d N=3 CSM quivers}}
To illustrate the conceptual difference of $\Ncal=3$ CSM theories compared to $\Ncal=4$ CSM theories, consider the following example: $M$ many NS5 branes on the left-hand-side, $P$ many $(1,\kappa_1)$ 5-branes in the middle, and $M$ many $(1,\kappa_2)$ 5-branes on right-hand-side; recall $0<\kappa_1<\kappa_2$. All of them are intersected by D3-branes in the following pattern.
\begin{align}
    \raisebox{-.5\height}{\includegraphics[width=1\linewidth]{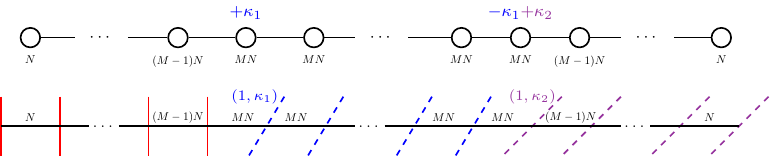}}
\end{align}
The non-trivial CS-levels of the quiver are $k_1=\kappa_1>0$ and $k_2 =\kappa_2 -\kappa_1>0$.
Based on the brane system in the factorised phase, there are three maximal branches, each of which with a single non-abelian isometry: 
\begin{align}
    \ffrak_{\mathrm{NS}} =\surmL(M)_{\mathrm{NS}} 
    \;, \quad 
    \ffrak_{(1,\kappa_1)} =\surmL(P)_{(1,\kappa_1)}
    \;, \quad
    \ffrak_{(1,\kappa_2)} =\surmL(M)_{(1,\kappa_2)} 
    \;.
\end{align}
In addition, there exists the $\urm(1)_\Ocal$ 0-form symmetry with its basic charged operator $\Ocal$:
\begin{align}
\label{eq:N=3_ex_monopole_ex1}
  \mono{(\bullet,\ldots,\bullet,(k_2/g,\circ),\bullet,\ldots,\bullet,(\circ,-k_1/g),\bullet,\ldots,\bullet)}  
  \;, \quad g=\gcd(\kappa_1,\kappa_2)=\gcd(k_1,k_2)\,,
\end{align}
having trivial magnetic flux at every non-CS node, flux $(k_2/g,\circ)$ at the $\urm(M\, N)_{k_1}$ node and flux $(\circ,-k_1/g)$ at the $\urm(M\, N)_{k_2}$ node. 

\paragraph{Faithful 0-form symmetry.}
The monopole operator~\eqref{eq:N=3_ex_monopole_ex1} transforms as
\begin{align}
    \left( [0,\ldots,0,\tfrac{k_2}{g}]_{\mathrm{NS}}\times [\tfrac{k_2+k_1}{g},0,\ldots,0]_{(1,\kappa_1)} \times [0,\ldots,0,\tfrac{k_1}{g}]_{(1,\kappa_2)}  \right)^{[q=1]}
\end{align}
under the group $\surm(M)_{\mathrm{NS}}\times \surm(P)_{(1,\kappa_1)}\times \surm(M)_{(1,\kappa_2)}\times \urm(1)_\Ocal$, wherein $q$ denotes the $\urm(1)_\Ocal$ charge. In other words, the centre charges of $\Ocal$ are
\begin{align}
\label{eq:N=3_ex_charges_Ocal_ex1}
    \Ocal: \quad
    \Big(-\tfrac{k_2}{g} ,\; \tfrac{k_2+k_1}{g}, \; -\tfrac{k_1}{g}; \; 1 \Big) \in \Z_M \oplus \Z_P \oplus \Z_M \oplus \Z
\end{align}
and any discrete charge can be suitably compensated by a $\urm(1)$ phase. Concretely, define
\begin{align}
    \label{eq:compensating_U1-Ocal_ex1}
    x_L= \left(\zeta_M , 1,1; e^{\tfrac{2\pi \im k_2}{g M}} \right), \,\,
     x_P= \left(1, \zeta_P ,1; e^{-\tfrac{2\pi \im (k_2+k_1)}{g P}} \right) ,\,\,
      x_R= \left(1 , 1,\zeta_M; e^{\tfrac{2\pi \im k_1}{g M}} \right) ,
\end{align}
with $\zeta_M$ (resp.\ $\zeta_P$) the standard generators for $\Z_M$ (resp.\ $\Z_P$); $x_{L,R}$ refer to the left and right $\Z_M$ factors in \eqref{eq:N=3_ex_charges_Ocal_ex1} and $x_P$ to $\Z_P$.
This implies that the faithful symmetry group is
\begin{align}
    \Fcal =  \frac{\surm(M)_{\mathrm{NS}}\times \surm(P)_{(1,\kappa_1)}\times\surm(M)_{(1,\kappa_2)}\times \urm(1)_{\Ocal}}{\langle  x_L, x_P, x_R\rangle} \, .
\end{align}

\paragraph{Non-genuine operators and extended group.}
The two guiding types of non-genuine monopole operators, see \eqref{eq:N=3_ex_mono-probes}, are
\begin{align}
    \label{eq:N=3_ex_mono-probes_ex1}
    \mono{\mathrm{probe},1} \equiv \mono{(\bullet,\ldots,\bullet,(1,\circ),\bullet,\ldots,\bullet,(0,\circ),\bullet,\ldots,\bullet)} \,, \quad
     \mono{\mathrm{probe},2} \equiv \mono{(\bullet,\ldots,\bullet,(0,\circ),\bullet,\ldots,\bullet,(1,\circ),\bullet,\ldots,\bullet)} \,.
\end{align} 
Inspecting the associated centre charges 
\begin{align}
\label{eq:N=3_ex_mono-probes_charges}
    \mono{\mathrm{probe},1} : 
    (-1,1,0;0) 
    \qquad
    \text{and}
    \qquad 
    \mono{\mathrm{probe},2} : 
    (0,-1,1;0)
\end{align}
shows that the group leaving $\mono{\mathrm{probe},i}$ as well as $\Ocal$ invariant is $\Z_d^{\mathrm{diag}} \subset \Z_M \times \Z_P \times \Z_M$ with $d=\gcd(M,P)$. Therefore, the extended 0-form group is
\begin{align}
    F=  \frac{\surm(M)_{\mathrm{NS}}\times \surm(P)_{(1,\kappa_1)}\times\surm(M)_{(1,\kappa_2)}}{\Z_d^{\mathrm{diag}}} \times \urm(1)_{\Ocal}
\end{align}
In addition, the relation between $\Fcal$ and $F$ is 
\begin{align}
    \Fcal \cong \frac{F}{\Z_{\ell} \times \Z_M} \qquad \text{with} \qquad 
    \ell = \lcm{M}{P}\;. 
\end{align}
To see this, define $\bar{M} = \frac{M}{d}$ and $\bar{P} = \frac{P}{d}$. The $\Z_d^{\mathrm{diag}}$ is generated by $\zeta_d=\big(\zeta_M^{\bar{M}}, \zeta_P^{\bar{P}}, \zeta_M^{\bar{M}}; 1 \big)$, such that 
\begin{align}
    \Zcal = \frac{\Z_M \times \Z_P \times \Z_M \times \urm(1)}{\langle \zeta_d \rangle} \,.
\end{align}
To define the generators of  $\Z_\ell \times \Z_M$, choose B\'ezout integers $r,s$ such that $\bar{P} r - \bar{M} s=1$. The two central elements of $F$ are
\begin{align}
    \gamma_\ell = \left(\zeta_M^r,\zeta_P^s,\zeta_M^{r}; e^{\tfrac{2\pi \im (k_2-k_1)}{g \ell}} \right) \;, \qquad 
    \gamma_M = \left(1,1,\zeta_M^{-1}; e^{\tfrac{2\pi \im k_1}{g M}} \right) .
\end{align}

\paragraph{2-group.}
To determine the 2-group structure, proceed as before by using lines. 
The first ingredient is to understand the 1-form symmetry $\Z_g^{(1)}$, with $g=\gcd(\kappa_1, \kappa_2)$; specifically, one needs to show
\begin{subequations}
    \begin{alignat}{3}
    W_1 &\coloneqq W_{F \text{ of } \urm(MN)_{k_1}}   &\qquad &\text{s.t.} \quad  & (W_1)^{\otimes g} &\sim \mathds 1\\
     W_2 &\coloneqq W_{F \text{ of } \urm(MN)_{k_2}}  & \qquad &\text{s.t.} \quad  & (W_2)^{\otimes g} &\sim \mathds 1
    \end{alignat}
\end{subequations}
of course, the bifundamental matter content already identifies $W_1 \sim W_2$ via local operator junctions.
In terms of charge cancellation, suppose that $(W_1)^{\otimes g}$ can end on a bare monopole
\begin{align}
   \mono{(m^{(M)},m^{(M+P)})}\coloneqq  \mono{(\bullet,\ldots,\bullet,m^{(M)},\bullet,\ldots,\bullet,m^{(M+P)},\bullet,\ldots,\bullet)}.
\end{align}
The overall $\urm(1)$ (gauge) charges at the two CS nodes of $(W_1)^{\otimes g}\ \mono{(m^{(M)},m^{(M+P)})}$ are
\begin{align}
   \Big( g + k_1 \sum_{i} m_i^{(M)} , \; k_2 \sum_{j} m_j^{(M+P)} \Big)
\end{align}
A necessary condition for being dressable by a chain of bifundamentals between node $\urm(M N)_{k_1}$ and $\urm(M N)_{k_2}$ is that the electric charges are opposite to one another, \ie
\begin{align}
    g + k_1 \sum_{i} m_i^{(M)} = - k_2 \sum_{j} m_j^{(M+P)}
\end{align}
writing $k_i = g \tilde{k}_i$ with $\gcd(\tilde{k}_1,\tilde{k}_2)=1$, this becomes
\begin{align}
     1 =  - \tilde{k}_1 \sum_{i} m_i^{(M)}  -  \tilde{k}_2 \sum_{j} m_j^{(M+P)} \;. \label{eq:1-form_N=3_CSM_ex}
\end{align}
This is exactly an equation of B\'ezout lemma type: for coprime $\tilde{k}_1, \tilde{k}_2$ there exist integers $p,q$ such that $\tilde{k}_1 p + \tilde{k}_2 q =1$. Therefore, a solution to \eqref{eq:1-form_N=3_CSM_ex} exists: 
\begin{align}
   \sum_{i} m_i^{(M)} =-p + \tilde{k}_2 t \quad \text{and} \quad \sum_{j} m_j^{(M+P)} = -q - \tilde{k}_1 t 
   \qquad \text{for any } t\in\Z \;.
\end{align}
To determine the 2-group, the centre charges of $\mono{(m^{(M)},m^{(M+P)})}$ are computed:
\begin{align}
    (p - \tilde{k}_2 t,q  -p + (\tilde{k}_2 +\tilde{k}_1) t,-q-\tilde{k}_1 t;t)
    \,.
\end{align}
One observes that they are $\Z_{d}^{\mathrm{diag}}$ invariant. Therefore, the endpoint $\mono{(m^{(M)},m^{(M+P)})}$ of $(W_1)^{\otimes g}$ transforms in $F$, as it should be by definition.
Moreover, the $t$-dependent part is not relevant, as this corresponds to a genuine operator transforming in $\Fcal$, see \eqref{eq:N=3_ex_charges_Ocal_ex1}. One is left with $(p,-p+q,-q;0)$, and the 2-group is defined by the short exact sequence
\begin{align}
    0 \to \Gamma^{(1)} = \Z_g^{(1)} \to \Ecal \to \Z_\ell \times \Z_M \to 0
\end{align}
which splits whenever $\gcd(g,M)=1=\gcd(g,P)$.

\paragraph{Anomalies.}
The anomalies involving the 1-form symmetry are derived as detailed in Section~\ref{subsec:sym_structure_N=3}. Concretely, the 1-form self-anomaly has anomaly label
\begin{align}
    p_{\gcd} =( \tilde{k}_1 +\tilde{k}_2) M N
\end{align}
and one can restrict to the anomaly-free subgroup $\Z_h^{(1)}\subset \Z_g^{(1)}$, see~\eqref{eq:anomaly-free_1-form}.
In the case of a split 2-group, the mixed 0-form and 1-form anomalies are all trivial.

\subsection{Remarks}
In this section, charge conjugation symmetry and non-Lagrangian 3d $\Ncal =3$ CSM theories are commented on.

\subsubsection{Charge conjugation}
\label{subsubsec:charge_conj}
The algorithm developed to study 2-group symmetries can be generalised to include the $\Z_2^C$ charge conjugation 0-form symmetry. The Wilson lines transform under $\Z_2^C$, presenting non-trivial 1-charges for the 0-form symmetry in the sense of \cite{Bhardwaj:2023wzd}. This implies that a 2-group involving the charge conjugation is characterised by a non-trivial action $\rho$ of the 0-form symmetry on the 1-form symmetry:
\begin{equation}
    \rho: \Z_2^C \rightarrow \text{Aut}(\Gamma^{(1)}) \, .
\end{equation}

Charge conjugation acts also on the 0-form symmetry group $F$ as an outer-automorphism, sending genuine and non-genuine operators that transform in the fundamental representation of $F$ to operators in the anti-fundamental one. 
The 2-group analysis presented in the previous section can be carried over, with some subtleties in the formalism adopted.
Considering, for concreteness, the 3d $\Ncal=4$ examples presented in Section~\ref{sec:examples_N=4}, once the charge conjugation symmetry is included, the novelties are:
\begin{itemize}
    \item In Examples 1 and 2, the 2-group structure becomes non-trivial, it can be denoted as $\Gamma^{(1)}\rtimes_\rho \Z_2^C$. Moreover, the mixed anomalies between the 1-form symmetry and the $\urm(1)$ 0-form symmetries need to be rephrased into the mixed anomaly between the 2-group and the $\urm(1)$ symmetry.
    \item In Example 3, 4 and 5 the non-trivial 2-groups are enriched by the $\Z_2^C$ symmetry. Given the non-trivial action of $\Z_2^C$ on $\Zcal$ and $\Gamma^{(1)}$, the 2-group is characterised by the twisted Postnikov class:
    \begin{equation}
        \widetilde{\beta} \in H_{\omega_1}^3(B(\Fcal \rtimes \Z_2^C), \Gamma^{(1)}) \, .
    \end{equation} 
    The cohomology group is twisted by the class $\omega_1 \in H^1(B(\Fcal \rtimes \Z_2^C),\Z_2^C)$ defined through the action $\rho$ and related to the background field of the charge conjugation symmetry, see \cite{Bhardwaj:2022scy, Hsin:2020nts}. $\widetilde{\beta}$ is defined from the original $\omega_2$ obstruction class of $\Fcal$ using the $\omega_1$-twisted Bockstein homomorphism in the long exact sequence for $\omega_1$-twisted cohomologies associated to the usual short exact sequence \eqref{shortexact}.
    Consequently, the constraint \eqref{eq:setup_1-form_background} on the 1-form background is upgraded with $\widetilde{\beta}$ and the definition of the 2-group background field can also be worked out.
\end{itemize}

A similar analysis can be carried out for 3d $\Ncal=3$ theories examined in Section~\ref{subsec:sym_structure_N=3}.
A more detailed analysis of the disconnected 2-group structure in the presence of the charge conjugation symmetry in the class of $\Ncal \geq 3$ CSM theories is open for future exploration.

\subsubsection{\texorpdfstring{Non-Lagrangian 3d $\Ncal=3$ CSM theories}{Non-Lagrangian 3d CSM theories}}
If one allows for various $(p_i,q_i)$ 5-branes with $p_i>1$, the resulting $\Ncal=3$ CSM theory is non-Lagrangian and can be written using sequences of $\Scal$ and $\Tcal$ duality-walls \cite{Assel:2014awa}.
Nevertheless, the features highlighted for the Lagrangian $\Ncal=3$ theories hold here as well. 

\paragraph{1-form symmetry.}
In the presence of NS5 branes and various $(p_i,q_i)$ 5-branes, the 1-form symmetry is $\Z^{(1)}_{g}$ with $g=\gcd(q_1,q_2,\dots)$. To see that, recall the quotient of the $(p,q)$ string lattice 
\begin{align}
    \Z^2 \slash \langle (1,0),(p_1,q_1) ,(p_2,q_2) , \ldots \rangle \cong \Z_{g}\,.
\end{align}
This generalises straightforwardly to a system of $(p_1,q_1)$ and $(p_2,q_2)$ 5-branes (without an NS5), for which the 1-form symmetry becomes
\begin{align}
      \Z^2 \slash \langle (p_1,q_1),(p_2,q_2) \rangle \cong \Z_{\Delta}\,,
      \quad \text{with } 
      \Delta = |\det{ 
      \begin{smallmatrix}
          p_1 & p_2\\
          q_1 & q_2
      \end{smallmatrix}}
      | = |p_1 q_2 - p_2 q_1|
      \,.
\end{align}
This brane-based argument is, however, not sensitive to the 1-form self anomaly.

\paragraph{0-form symmetry and 2-group structure.}
Since one can bring the system in its factorised phase even in the non-Lagrangian case, it is possible to obtain $\Fcal_{(p,q)}$ (see \cite{Marino:2025uub}).
However, it does not seem feasible to study either $F$ or the 2-group structure, since the theory is non-Lagrangian and the end-point operators cannot be written down clearly.
\section{\texorpdfstring{Maximal branches and MQs of 3d $\Ncal\geq3$ CSM theories}{Max branches and MQs for 3d CSM theories}}
\label{sec:maxbranch_MQ}
Having analysed the symmetry structures of $\Ncal \geq 3$ CSM quivers, the next step is to probe how much of this affects the moduli space of vacua. Concretely, parts of the 0-form symmetry are hyper-K\"ahler isometries of maximal branches. However, as fractional monopole operators are endpoints of 1-form symmetry lines, maximal branches may or may not be enlarged once such fractional monopoles become genuine operators.

\subsection{From frozen D3 branes to ``frozen'' TQFTs}
\label{subsec:frozen_TQFTs}
Starting from the Type IIB brane configuration, a maximal branch is best analysed via its \emph{maximal branch phase}, as defined in Section~\ref{subsec:UnitaryCSM_setup}.
A maximal branch is parametrised by a subset of all gauge-invariant dressed monopole operators. The distinguishing feature of a maximal branch dressed monopole operator is that the fields used to dress the (generically non-gauge invariant) bare monopole $\mono{m}$ operator remain massless in the monopole background $m$, \ie they belong to the residual massless theory, and that they stem exclusively from hypermultiplets or twisted hypermultiplets.\\

Compared to the Type IIB brane setup of NS5, D5, and D3 branes that give rise to non-CS 3d $\Ncal=4$ theories, the CSM brane system differs in an important aspect: brane creation and annihilation between $(1,\kappa_i)$ and $(1,\kappa_j)$ 5-brane proceeds in units of $\abs{\kappa_j-\kappa_i}$ D3s (see GK duality in Section~\ref{subsec:UnitaryCSM_setup}), whereas brane creation and annihilation between NS5 and D5 proceeds in $1$ unit of D3. Therefore, the maximal branch phase in an NS5-D5-D3 system never has any frozen D3 brane\footnote{Recall that a D3 in between two different types of $(p,q)$ 5-branes cannot move along any of the transverse directions. This is referred to as ``frozen''.}. In contrast, a maximal branch phase in an $(1,\kappa_1)$-$(1,\kappa_2)$-$\dots$-D3 system may exhibit a number of frozen D3s that cannot be eliminated by brane creation or annihilation moves. These frozen D3s do not affect the maximal branch geometry, see \cite{Marino:2025uub}, as long as the 1-form symmetry is not gauged. In this section, the aim is to explore how the maximal branches are affected by 1-form symmetry gauging.

To set the stage, consider a maximal branch phase in which there are $\ell$ D3 branes frozen in between a $(1,\kappa_1)$ and a $(1,\kappa_2)$ 5-branes.
\begin{equation}
    \raisebox{-.5\height}{\includegraphics[scale=1.2]{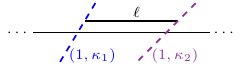}}
    \label{eq:frozenD3s}
\end{equation}
Now define $k_{ij}=\kappa_i-\kappa_j$.
Since the CS coupling gives a mass to the adjoint chiral multiplet of the $\Ncal=4$ vector multiplet, the world-volume of the frozen D3s carries a 3d $\Ncal=2$ pure $\urm(\ell)_{k_{21}}$ CS theory. Upon exchanging the $(1,\kappa_1)$ and the $(1,\kappa_2)$ 5-branes, the dual brane system exhibits $\abs{k_{21}}-\ell$ many frozen D3s, whose world-volume theory is a 3d $\Ncal=2$ pure $\urm(\abs{k_{21}}-\ell)_{k_{12}}$ CS theory.

The fate of such a $\Ncal=2$ pure CS theory in the IR is well-known: upon integrating out massive gauginos the CS-level of the non-abelian factor shifts, while the abelian level does not \cite{Intriligator:2013lca}. This results in an IR TQFT of the following type:
\begin{alignat}{3}
\label{eq:frozen_TQFT}
\begin{aligned}
    \Ncal=2\; &\urm(\ell)_{k_{21}} \quad & &\longrightarrow \quad & T_\ell &\coloneqq \urm(\ell)_{\abs{k_{21}}-\ell,\,k_{21}} \,, \\
    \Ncal=2 \; &\urm(\abs{k_{21}}-\ell)_{k_{12}} \quad &  &\longrightarrow \quad & T^\vee_\ell &\coloneqq  \urm(\abs{k_{21}}-\ell)_{-\ell,\,k_{12}}
    \,,
\end{aligned}
\end{alignat}
with the definition $\urm(\ell)_{k-\ell,\,k} \coloneqq (\surm(\ell)_{k-\ell} \times \urm(1)_{\ell k})\slash \Z_\ell$.
Note that the two TQFTs are level/rank dual, $T_\ell= \urm(\ell)_{\abs{k_{21}}-\ell,\,k_{21}}\leftrightarrow \urm(\abs{k_{21}}-\ell)_{-\ell,\,k_{12}} =T^\vee_\ell$, cf.~\cite{Hsin:2016blu,Delmastro:2019vnj}. This is not surprising, since they originate from GK-dual CSM theories. In particular, for $\ell\in\{0,\abs{k_{21}}\}$ the same duality certifies that $T_\ell$ is trivial, since $\urm(\abs{k_{21}})_{0,\,k_{21}}\leftrightarrow \urm(0)_{k_{12},\,k_{12}}$.

The Chern--Simons TQFT has a variety of Wilson lines. A Wilson line of $\urm(\ell)_{k_{21}-\ell,\,k_{21}}$ is labelled by integer weights $\lambda=(\lambda_1,\ldots,\lambda_{\ell-1})$ of $\surm(\ell)$ together with a $\urm(1)$ charge $q$, subject to the $\Z_\ell$ quotient constraint $q\equiv |\lambda| \bmod \ell$. Consider the non-trivial Wilson line 
\begin{align}
\label{eq:det-Wilson_TQFT}
U\equiv W_{(\lambda=0,\,q=\ell)} \equiv W_{\mathrm{det}}  = \exp(\im \oint \sum_{h=1}^\ell c_h) \;,
\end{align}
which is the Wilson line in the determinant representation of $\urm(\ell)$ and the $c_h$ denote the $\ell$ Cartan components of the $\urm(\ell)$ gauge field. 
The line $U$ cannot end. In fact, none of the $U^{\otimes a}$ with $a\in\{1,2,\ldots,\abs{k_{21}}-1\}$ can end, since $U^{\otimes a}$ carries $\urm(1)$ charge $a\ell$, which is non-trivial in $\urm(1)_{\ell k_{21}}$. However, $U^{\otimes \abs{k_{21}}}\sim 1$ is equivalent to the trivial line, because $\abs{k_{21}}\ell\equiv 0 \bmod \ell\abs{k_{21}}$ --- in other words, $U^{\otimes\abs{k_{21}}}$ can end on the unit monopole operator.
This establishes a $\Z_{\abs{k_{21}}}^{(1)}$ symmetry in the TQFT, realised by the anyonic Wilson line $U$. Associated to anyons is the topological spin~\cite{Hsin:2018vcg}, which reads
\begin{align}
\label{eq:top_spin_TQFT}
    h[U^{\otimes a}] = \frac{q(U^{\otimes a})^2}{2\abs{k_{21}} \ell} = \frac{\ell}{2\abs{k_{21}}} a^2 \;, \qquad a\in\{0,1,\ldots,\abs{k_{21}}-1\} \,.
\end{align}
Hence $T_\ell$ carries the anomaly label $p_{\mathrm{TQFT}}=\ell$. The obstruction to gauging is not the non-integrality of $h$ as such: in a spin theory, a fermionic line with $h\in\tfrac12+\Z$ is still condensable, and only $2h\notin\Z$ obstructs. Equivalently, the $\Z_g^{(1)}$ subgroup generated by $U^{\otimes \abs{k_{21}}/g}$ is anomaly-free precisely because
\begin{align}
\label{eq:1-form_subgroup_TQFT}
    h[U^{\otimes \abs{k_{21}}/g}] = \frac{\ell \abs{k_{21}}}{2 g^2} \in \tfrac12 \Z
    \qquad \text{for } g=\gcd(\ell,\abs{k_{21}}) \,.
\end{align}

Consequently, if a maximal branch phase has frozen D3s in it, there exists at least one Chern--Simons TQFT that exhibits the same 1-form symmetry as the CSM theory.
Whether or not a maximal branch is affected by 1-form symmetry gauging crucially depends on whether or not a TQFT is present, as discussed below.

\subsection{Maximal branches and 1-form symmetry gauging}
\label{subsec:max_branch_1form_gauging}
Consider a linear 3d $\Ncal\geq3$ CSM quiver theory realised by a $(1,\kappa_1)$-$(1,\kappa_2)$-$\dots$-D3 brane system, endowed with a 1-form symmetry $\Z_g^{(1)}$ with anomaly-free subgroup $\Z_h^{(1)}$.
Focus on the $(1,\kappa_i)$ branch (assuming that there are at least two $(1,\kappa_i)$ 5-branes, in order for the associated branch to be non-trivial); \eg a configuration may locally look as follows.
\begin{align}
\raisebox{-.5\height}{
    \includegraphics[]{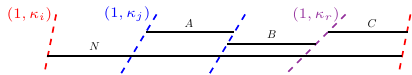}
    }
    \label{eq:Unitary_N3_brane_example1}
\end{align}
Then the $(1,\kappa_i)$ branch is affected by the 1-form symmetry gauging if and only if all the following requirements are met:
\begin{enumerate}[label=(MB\arabic*),ref=(MB\arabic*),leftmargin=*]
    \item \label{item:MaxBranch} 
        The system admits a $(1,\kappa_i)$ branch phase, in which there are no D3 moduli free to move along any other maximal branch. \ie no D3 segments that are free to move between two $(1,\kappa_j)$ branes for any $j\neq i$. \eg $A=0$ in \eqref{eq:Unitary_N3_brane_example1}.
    \item \label{item:TQFT} There are no non-trivial frozen TQFTs, namely there are no frozen D3 branes stretched between two 5-branes of different type. \ie no D3 segments between $(1,\kappa_r)$ and $(1,\kappa_s)$ for any $r\neq s$. \eg $B=0$ and $C=0$ in \eqref{eq:Unitary_N3_brane_example1}.
\end{enumerate}
Concretely, this translates into the requirement that the $(1,\kappa_i)$ branch phase satisfies: 
\begin{equation}
    N_i\equiv N \quad\forall i
    \quad \text{and} \quad
    \sum_{i} k_i =0 
    \,.
    \label{eq:constraint_maxbranch_affected}
\end{equation}
\ie in between any two adjacent $(1,\kappa_i)$ 5-branes, the number of D3s is constant regardless of how many $(1,\kappa_j)$ 5-branes $j\neq i$ are in between, and the sum of CS-levels is zero.\\

Condition \ref{item:MaxBranch} is really just the definition of the maximal branch phase for the $(1,\kappa_i)$ 5-branes. In contrast, condition \ref{item:TQFT} guarantees that the maximal branch is affected by 1-form symmetry gauging. In other words, if condition \ref{item:TQFT} holds, then the end point of the 1-form generator $D$ is a monopole operator $\mono{\mathrm{frac}}$ with fractional magnetic fluxes that lives on the $(1,\kappa_i)$ maximal branch. \\

To see that these conditions are necessary and sufficient, focus on a single $(1,\kappa_i)$ 5-brane interval, with $N$ D3 maximal branch moduli, some $n_r$ frozen D3s, and vanishing sum of CS-levels. For instance, consider the following case with $\ell-1$ many potentially different $(1,\kappa_j)$ 5-branes in between two $(1,\kappa_i)$ 5-branes.
\begin{align}
\raisebox{-.5\height}{
    \includegraphics[]{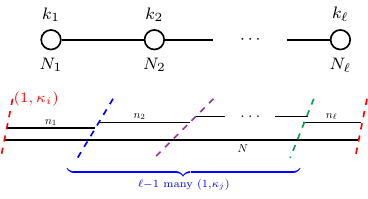}
    }
    \label{eq:Unitary_N3_brane_example2}
\end{align}
At a generic point of the $(1,\kappa_i)$ branch, the $N$ D3 moduli in between the $(1,\kappa_i)$ are separated in the transverse direction. Thus, the $\urm(N)$ gauge group in the D3 world volume is broken to the maximal torus $\urm(1)^N$. 

For each $\urm(N+n_r)$ gauge field $A^{(r)}$ in \eqref{eq:Unitary_N3_brane_example2}, split into Cartan components $A_i^{(r)}$ for the moduli directions $i=1,\ldots,N$ and the pure CS directions $A_{N+t}^{(r)}=c_t^{(r)}$ for $t=1,\ldots, n_r$, cf.\ \eqref{eq:det-Wilson_TQFT}. For a fixed moduli direction $i$, perform a uni-modular transformation defined via
\begin{align}
\label{eq:uni-modular_trafo}
\begin{cases}
       b_i^{(r)} &= A_i^{(r)} - A_i^{(r+1)} \;, \text{for } r =1,2,\ldots, \ell-1\;,\\
    a_i &= A_i^{(\ell)} \,.
\end{cases}
\end{align}
In this basis, generically all $b_i^{(r)}$ have a CS-term as well as a BF-coupling with $a_i$; whereas, $a_i$ itself has no CS-term. The $a_i $ are the moduli fields. The other $b_i$ fields all trigger a BF-reduction, which reproduce the expected 1-form symmetry in this 5-brane configuration. The action of the centre fractional defect $D$ for the anomaly-free $\Z_h^{(1)}\subset \Z_g^{(1)}$ on the moduli directions reduces to 
\begin{align}
    \alpha_\Sigma=(\alpha_{b^{(1)}},\alpha_{b^{(2)}},\ldots,\alpha_{b^{(\ell-1)}},\alpha_a) = (0,0,\ldots,0, \frac{1}{h}\mathds{1}_N)
\end{align}
\ie only the $a_i$ perceive the monodromy. This allows one to deduce the fate $D \to D_\Sigma$ of $D$ in the sigma model, because the topological spin of $D_\Sigma$ (defined by $\alpha_\Sigma$) is evaluated via
\begin{subequations}
\begin{align}
    h[D_\Sigma] &= \frac{1}{2} \alpha_\Sigma^T \tilde{K} \alpha_\Sigma = 
    \frac{1}{2} \alpha_M^T K \alpha_M = 
    \frac{1}{2h^2} \sum_{r=1}^\ell k_r N =0
    \\
    \text{with }
    \alpha_M&= (\alpha_{A^{(1)}},\alpha_{A^{(2)}},\ldots,\alpha_{A^{(\ell-1)}},\alpha_{A^{(\ell)}}) = \frac{1}{h} (\mathds{1}_N,\mathds{1}_N,\ldots,\mathds{1}_N,\mathds{1}_N) \;,
\end{align}    
\end{subequations}
where $\alpha_M$ are the monodromies along the moduli directions of the $A^{(r)}$ gauge fields. Here $K =\diag(k_1,k_2,\ldots,k_\ell)$ the matrix of CS-levels, and $\tilde{K}$ is the matrix obtained after unimodular transformation~\eqref{eq:uni-modular_trafo}. Therefore, the anomaly label $p_\Sigma$ of $D_\Sigma$ is trivial, and $D_\Sigma$ can be seen as a vortex line $V_{\alpha_\Sigma}$ purely in the moduli fields $a_i$.

The frozen $n_r$ D3s each define a ``frozen'' TQFT $T_{n_r}$, see Section~\ref{subsec:frozen_TQFTs}. Since the anomaly label $p_\Sigma$ of the maximal branch sigma-model is trivial, the TQFTs need to satisfy the IR anomaly matching condition
\begin{align}
    p_{\mathrm{UV}} = \sum_{i=1}^{\ell} N_i \frac{k_i}{g}  = \underbrace{\sum_{i=1}^{\ell} N \frac{k_i}{g}}_{\equiv p_\Sigma=0} + \sum_{i=1}^{\ell} n_i \frac{k_i}{g}\stackrel{!}{=}    p_{\mathrm{IR}} = \sum_r p_{T_{n_r}} 
\end{align}
where $N_i= N + n_i$ denotes the number of D3s in each 5-brane interval.
Restrict to the anomaly-free 1-form symmetry $\Z_h^{(1)}$, for which $p_{\mathrm{UV}}=0$, then necessarily all frozen TQFTs $T_{n_r}$ have $\Z_h^{(1)} \subset \Gamma^{(1)}_{T_{n_r}}$. This follows simply because 
\begin{align}
\Gamma^{(1)}_{T_{n_r}} = \Z_{k_{i,j}^{(r)}}^{(1)}\quad   \text{ for some } k_{i,j}^{(r)} =| \kappa_i-\kappa_j| \text{ with } i\neq j \;,
\end{align}
and $\Z_h^{(1)}\subset \Z_g^{(1)}$ is defined by $g=\gcd(\kappa_1,\kappa_2,\ldots)$ and $h\mid g$; hence, $h \mid k_{i,j}^{(r)}$ for all r. Therefore, the centre fractional Gukov--Witten defect $D$ is diagonally embedded\footnote{Alternatively, recall the definition of the centre fractional defect~\eqref{eq:diag_centre_defect}, which implies that all the pure $\urm(n_r)$ CS theories experience the same monodromy simultaneously.} as
\begin{align}
    D \to D_\Sigma  \times \prod_{r} U_r^{\otimes\, k_{i,j}^{(r)}/h} \,,
\end{align}
where $U_r$ denotes the 1-form generator of $T_{n_r}$, see \eqref{eq:det-Wilson_TQFT} and \eqref{eq:1-form_subgroup_TQFT}. However, any such anyonic Wilson line has no endpoint. Therefore, the presence of any non-trivial ``frozen'' TQFT prevents new fractional monopole operators from entering the maximal branch upon 1-form symmetry gauging.

If, on the other hand, all TQFTs are trivial, then $D \to D_\Sigma$, and the end-point is the fractional monopole operator $\mono{\mathrm{frac}}$ on which the vortex line $V_{\alpha_\Sigma}$ can end. Indeed, such a fractional monopole operator can be consistently dressed by fields that remain massless in the monopole background $\mono{\mathrm{frac}}$. To see this, it is sufficient to recall that the weights of a  $\urm(N_a)\times \urm(N_b)$ bifundamental hyper in the $(1,\kappa_i)$ 5-brane interval of \eqref{eq:Unitary_N3_brane_example2} is $ e_i^{(a)} - e_j^{(b)}$ with $i\in\{1,\ldots, N_a\}$, $j\in\{1,\ldots, N_b\}$, and $e^{(s)}_i$ fundamental weights of $\urm(N_s)$.
To determine massless-ness in the $\mono{\mathrm{frac}}$ background, one evaluates the pairing of the fractional magnetic flux on the weights:
\begin{align}
  \langle  (e_i^{(a)} - e_j^{(b)})  , \frac{1}{h}((1,\ldots,1),(1,\ldots,1))\rangle =0 \quad \text{for any }i,j \,.
\end{align}
Thus, the endpoint of $D$ is not only non-empty, but it is in fact a dressed maximal branch monopole operator. See Appendix~\ref{app:dressed_monopoles} for further details.

\paragraph{How many maximal branches can be affected?}
Having established the criteria under which a 1-form symmetry gauging affects a chosen maximal branch, the last question to be addressed is how many maximal branches can be affected simultaneously. Consider, without loss of generality, a system with NS5, $(1,\kappa_1)$, $(1,\kappa_2)$ and possibly more 5-branes, arranged as follows:
\begin{equation}
    \raisebox{-.5\height}{\includegraphics[width=0.55\linewidth]{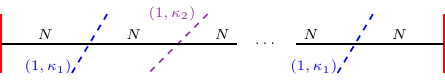}}
    \label{eq:Unitary_N3_1kbranch_1}
\end{equation}
\begin{itemize}
    \item The system is already in the NS5 branch phase, and since it satisfies the constraints \eqref{eq:constraint_maxbranch_affected} one can readily conclude that the NS5 branch is affected by the $\Z_h^{(1)}$ 1-form symmetry gauging.
    \item To check whether the $(1,\kappa_1)$ branch is affected by the $\Z_h^{(1)}$ 1-form symmetry gauging, one has to bring the system \eqref{eq:Unitary_N3_1kbranch_1} in its $(1,\kappa_1)$ branch phase; to this end, move the two leftmost and rightmost $(1,\kappa_1)$ 5-branes through the NS5 branes on the left and on the right of the setup.
    The configuration after brane creation/annihilation is the following:
    \begin{equation}
        \raisebox{-.5\height}{\includegraphics[width=0.6\linewidth]{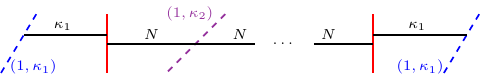}}
    \end{equation}
    Based on the constraints \eqref{eq:constraint_maxbranch_affected}, one readily concludes that the $(1,\kappa_1)$ branch is also affected by the $\Z_h^{(1)}$ 1-form symmetry gauging provided $\kappa_1=N$ holds.
    \item For the $(1,\kappa_2)$ branch to be affected by the $\Z_h^{(1)}$ 1-form symmetry gauging, one has to move two $(1,\kappa_2)$ 5-branes outside the NS5 interval into a maximal $(1,\kappa_2)$ branch phase with no frozen D3s (no non-trivial TQFT). However, since they are not near the brane system edges, one has to exchange at least once a $(1,\kappa_2)$ with a $(1,\kappa_1)$ via GK-duality. The result is a maximal branch phase with no frozen D3s if $\abs{\mathrm{det}\left(\begin{smallmatrix}1&1 \\ \kappa_1&\kappa_2\end{smallmatrix}\right)}=0$, namely if $\kappa_2=\kappa_1$.
    Since an iterative application of this reasoning leads to setting all other possible $\kappa_i=\kappa_1$, one concludes that no $(1,\kappa_{i\neq1})$ branch can be affected by the $\Z_h^{(1)}$ 1-form symmetry gauging.
\end{itemize}
This argument shows that gauging the $\Z_h^{(1)}$ 1-form symmetry can affect at most two maximal branches of a unitary $\Ncal\geq3$ CSM theory. For both branches to be affected, however, parameters such as $\kappa_i$ and $N$ must be highly constrained.

\subsection{Magnetic quivers and 1-form symmetry}
\label{subsec:MQ_1form}
The magnetic quivers for linear $\Ncal\geq3$ unitary CSM quivers were derived in \cite{Marino:2025uub}. Based on the discussion here, two questions are natural:
\begin{itemize}
    \item Can the magnetic quiver construction be extended to include 1-form symmetry?
    \item Can the magnetic quiver(s) capture (parts of) the 't~Hooft anomalies?
\end{itemize}
By the very idea of magnetic quivers, these are \emph{auxiliary} objects whose 3d $\Ncal=4$ Coulomb branch is an honest description of a maximal branch (i.e.\ the target of the IR sigma model). It is therefore not surprising that the encountered frozen TQFTs (and their effect on the maximal branch being affected by 1-form symmetry gauging or not) are not part of the pure quiver data imprinted on the magnetic quiver.

\subsubsection{Magnetic quiver extension: non-simply laced edges}

\paragraph{$\MQNS$ extension.}
Consider the following setup, already in the NS5 branch phase and with no frozen TQFTs, see conditions \ref{item:MaxBranch} and \ref{item:TQFT}.
\begin{equation}
    \raisebox{-.5\height}{\includegraphics{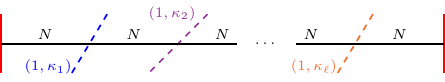}}
    \label{eq:Unitary_N3_ExampleTheoryForMQ}
\end{equation}
The magnetic quiver extension for the NS5 branch (\ie \(\MQNS\)) is \emph{conjectured} to be
\begin{equation}
    \raisebox{-.5\height}{\includegraphics[width=0.35\linewidth]{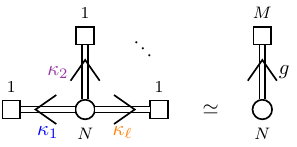}}
    \label{eq:nNodes_MQ_nonSimplyLaced_1node}
    \;, \quad \text{with }
    \begin{cases}
        M=\sum_{i=1}^{\ell} \frac{\kappa_i}{g} \;, \\[5pt]
        g=\gcd(\kappa_1,\kappa_2,\dots,\kappa_\ell) \,,
    \end{cases}
\end{equation}
and with a collection of \(\kappa_i\) non-simply laced edges, for $i=1,\dots,\ell$. 
Such edges have been introduced in \cite{Cremonesi:2014xha} and have since then appeared frequently in 3d $\Ncal=4$ Coulomb branch literature, especially in magnetic quivers \cite{Bourget:2020asf,Bourget:2020mez,Bourget:2022ehw}. Despite a lack of a Lagrangian interpretation, the Coulomb branch of quivers like \eqref{eq:nNodes_MQ_nonSimplyLaced_1node} can be probed via the monopole formula \cite{Cremonesi:2013lqa}.

The motivation for non-simply laced edges is the following. The \((1,\kappa_i)\) 5-brane is a single bound state with \(\kappa_i\) units of D5 charge. Hence, it is natural that such an object only contributes as a single flavour node in the magnetic quiver. The analogy follows from interpreting the position of the \((1,\kappa_i)\) 5-brane as a deformation parameter: thus, for a single such 5-brane, there should only be one deformation parameter, not \(\kappa_i\) many. (See also \cite{Santilli:2026gjy} for a discussion in the abelian case.) Of course, this does not affect the singular Coulomb branch of the magnetic quiver, but it becomes important once the 1-form symmetry, its gauging, and mixed anomalies are taken into account; cf. \cite{Nawata:2023rdx,Grimminger:2024doq}.

Concretely, as the NS5 branch of \eqref{eq:Unitary_N3_ExampleTheoryForMQ} is affected by gauging the 1-form symmetry all of the \(\kappa_i\) D5 charges of each $(1,\kappa_i)$ contribute to a single gauge node in the \(\MQNS\). In this circumstance, the new gauge-invariant dressed CSM monopole operators with fractional fluxes are obtained from the monopole operators with integer fluxes via an overall fractional shift. Having a collection of \(\kappa_i\) non-simply laced edge achieves the correct counting and, in fact, exhibits a \(\Z_g\) 1-form symmetry.\\

In general, one can consider a setup with multiple NS5 intervals.
\begin{equation}
    \raisebox{-.5\height}{\includegraphics{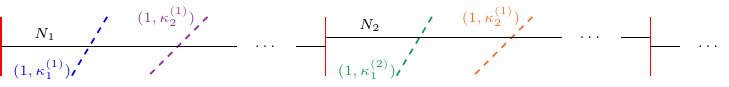}}
    \label{eq:Unitary_N3_ExampleTheoryForMQ_Nnodes}
\end{equation}
The $\MQNS$ extension then reads
\begin{equation}
    \raisebox{-.5\height}{\includegraphics[width=0.2\linewidth]{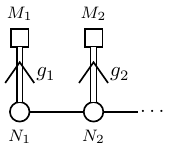}}
    \label{eq:nNodes_MQ_nonSimplyLaced_Nnode}
    \;, \quad \text{with}
    \begin{cases}
         M_{j}=\sum_{i=1}^{\ell_j} \frac{\kappa^{(j)}_i}{g_j} \;, \\[5pt]
         g_j=\gcd(\kappa^{(j)}_1,\kappa^{(j)}_2,\dots,\kappa^{(j)}_{\ell_j}) \,.
    \end{cases}
\end{equation}
and overall $\MQNS$ exhibits a $ \Z_g^{(1)}$ 1-form symmetry, with $g=\gcd(g_1,g_2,\ldots) = \gcd(\kappa_1,\kappa_2,\ldots) $. This matches the gaugeable 1-form symmetry of the CSM theory, because $p=0$ and $h=g$.

\begin{table}[!ht]
\centering
\begin{tabular}{ccc}
\toprule
    Example & A-branch & B-branch \\ \midrule
    1, Sec.~\ref{subsubsec:Ex1} & always affected & trivial \\
    & \raisebox{-.5\height}{\includegraphics[width=0.06\linewidth]{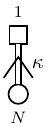}} & trivial \\
    \midrule
    2, Sec.~\ref{subsubsec:Ex2} & never affected & never affected \\
    \midrule
    3, Sec.~\ref{subsubsec:Ex3} & affected if $d=0$ and any $\kappa\geq N$ & affected if $\kappa=N$ and $d=0$ \\
    & \raisebox{-.5\height}{\includegraphics[width=0.06\linewidth]{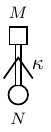}} &  
    \raisebox{-.5\height}{\includegraphics[width=0.3\linewidth]{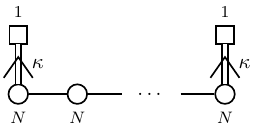}} \\
    \midrule
    4, Sec.~\ref{subsubsec:Ex4} & always affected & always affected \\
    & \multicolumn{2}{c}{
    \raisebox{-.5\height}{\includegraphics[width=0.3\linewidth]{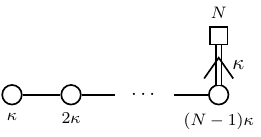}}
    }\\
    \midrule
    5, Sec.~\ref{subsubsec:Ex5} & affected for $LN<R\kappa$ & not affected for $LN<R\kappa$ \\
    & \multicolumn{2}{c}{$
    \raisebox{-.5\height}{\includegraphics[width=0.4\linewidth]{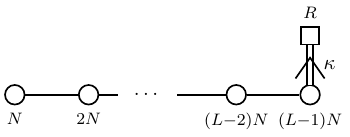}}$} \\[15pt]
    \cline{2-3} \\[-10pt]
    & \multicolumn{2}{c}{always affected for $LN=R \kappa$} \\
    & 
    $\raisebox{-.5\height}{\includegraphics[width=0.3\linewidth]{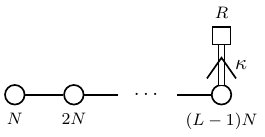}}$ &  
    $\raisebox{-.5\height}{\includegraphics[width=0.3\linewidth]{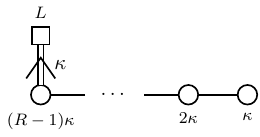}}$ \\
\bottomrule
\end{tabular}
\caption{Magnetic quiver extensions for the examples of Section~\ref{sec:examples_N=4}. Whenever the A or B branch is affected by 1-form gauging, the $\MQ$ exhibits a 1-form symmetry $\Z_\kappa^{(1)}$ and reproduces the maximal branch isometry as well as the mixed 't~Hooft anomaly between the 1-form symmetry and the isometry. Whenever no extension is required, the results of \cite{Marino:2025uub} apply directly.}
\label{tab:MQ_extension_N=4}
\end{table}

\paragraph{$\MQ_{\kappa_i}$ extension.}
The discussion readily generalises to any other $(1,\kappa_i)$ branch by dualising the system with the $\text{SL}(2,\Z)$ transformation $\Tcal^{\abs{\kappa_i}}$, see \cite{Marino:2025uub} for details. Bringing the dualised setup into its NS5 branch phase -- which is the $\Tcal^{\abs{\kappa_i}}$-dual of the $(1,\kappa_i)$ branch phase of the original theory -- and writing the magnetic quiver $\MQNS$ according to the logic explained above, one obtains the $\MQ_{\kappa_i}$ of the starting theory.

\paragraph{Mixed 't~Hooft anomalies.}
In general, the magnetic quiver extension $\MQ_{\kappa_i}$ captures
\begin{itemize}
    \item the $\Z_g^{(1)}$ 1-form symmetry, which is always gaugeable,
    \item the mixed 't~Hooft anomalies between Coulomb branch 0-form symmetry and $\Z_g^{(1)}$.
\end{itemize} 
To prove the claim of the mixed 0-form 1-form anomaly, it is sufficient to focus on a single NS5 interval in the maximal NS5 branch phase, \ie \eqref{eq:nNodes_MQ_nonSimplyLaced_1node}. Therein, the CSM quiver description gives rise to $\ell+1$ topological fugacities, say $w_i$. For the maximal branch, only the diagonal combination is relevant $a\sim \prod_i w_i$. Moreover, the maximal branch moduli in this NS5 segment are labelled by  $\urm(N)$ coweights\footnote{The magnetic fluxes of all $\ell+1$ nodes in the CSM quiver need to exactly coincide to yield maximal NS5 branch operators, analogous to what is shown in Appendix~\ref{app:dressed_monopoles} for an $\Ncal=4$ example.}; thus, the fractionally shifted coweights uniformly affect the NS5 branch symmetry fugacity $a$ via $a^{\sum_{j=1}^{N} m_j + \frac{N}{g}}$. Consequently, the mixed 0-form--1-form $ \frac{N}{g}B_2 \cup c_1(\urm(1)_a)$ coincide between CSM theory and magnetic quiver. 

\subsubsection{Examples of magnetic quiver extensions}
For the examples of Section~\ref{sec:examples_N=4}, the magnetic quiver extensions are summarised in Table~\ref{tab:MQ_extension_N=4}.
\section{Conclusion and outlook}
\label{sec:conclusion}
In this work two aspects of highly supersymmetric Chern--Simons matter theories realised as world-volume theories of Type IIB intersecting brane configurations have been studied: symmetries and their anomalies, and maximal branches and their interplay with higher symmetries.
The brane setup has been used in two complementary phases: in the factorised phase, the symmetry factorisation and the relevant monopole charges have been made manifest, while in the maximal branch phases the effect of 1-form symmetry gauging has been exposed.
A uniform treatment of the linear unitary $\Ncal=4$ and $\Ncal=3$ quiver theories realised by NS5 branes and $(1,\kappa_i)$ 5-branes has thereby been obtained.
In particular, higher-symmetry data have been connected to the geometry of maximal branches through line endpoints, monopole operators, and the frozen D3-brane sectors.

\paragraph{Symmetries and anomalies.}
The 1-form symmetry of a linear unitary CSM quiver has been determined to be $\Z_g^{(1)}$, with $g=\gcd(\{\kappa_i\})$, in two independent ways: from the screening of Wilson lines, and from the quotient of the $(p,q)$-string lattice by the strings that can end on the available 5-branes. Its generator has been realised as a fractional central Gukov--Witten defect, whose end-points are monopole operators with fractional magnetic fluxes. Demanding that such an operator admit a gauge-invariant dressing has produced both the self-anomaly label $p=\sum_i N_i k_i/g$, and the maximal anomaly-free (hence gaugeable) subgroup $\Z_h^{(1)}$ with $h=\gcd(g,p)$.
The 0-form global form has been extracted in the factorised phase from balanced and overbalanced nodes.
Balanced nodes are responsible for the non-abelian factors, and the associated monopole operators carry trivial centre charge; non-trivial centre charges are instead attributed to the unit monopoles at the overbalanced nodes, which determine the faithful group $\Fcal$, and to the non-genuine unit monopole at the CS-node, which probes the connected cover $F$.
The corresponding centre characters and probe charges are detailed in Section~\ref{subsec:general_N=4} for $\Ncal=4$ CSM quivers, and in Section~\ref{subsec:sym_structure_N=3} for $\Ncal=3$ CSM quivers.

For $\Ncal=4$ CSM quivers, a non-trivial 2-group involving $\Gamma^{(1)}$ and both $\Fcal_A$ and $\Fcal_B$ can be formed under the conditions demonstrated in Examples~4 and~5, in contrast to the non-CS 3d $\Ncal=4$ theories considered in~\cite{Bhardwaj:2023zix,Nawata:2023rdx}.
For $\Ncal=3$ CSM quivers, a potential non-trivial 2-group involving all non-abelian factors simultaneously has likewise been identified.

On the other hand, no non-trivial mixed 't~Hooft anomalies between the 1-form symmetry and non-abelian 0-form factors have been found in the linear unitary theories considered here; the cancellation has been tied both to the correct UV-to-IR symmetry map and to the balance condition, as made explicit around~\eqref{eq:balanced_nodes_no_mix_1-form}.

\paragraph{Maximal branches.}
The D3 segments frozen between adjacent 5-branes of different type support a pure $\Ncal=2$ Chern--Simons theory, which flows to a CS TQFT.
If a non-trivial such TQFT is present at a generic point of a maximal branch phase, that maximal branch is not affected by the gauging of the anomaly-free subgroup of the 1-form symmetry. 
This protection has been traced to the fact that the physical 1-form generator is the diagonal combination of a vortex line of the branch sigma model with the anyonic Wilson lines of the frozen TQFTs: the latter cannot be ended, hence no end-point is supported on the maximal branch.
Conversely, when all frozen TQFTs are trivial --- equivalently, when the gauge ranks are constant throughout the branch phase --- the generator reduces to a pure vortex line of the sigma model, whose end-point is a fractional-flux monopole operator that can be dressed by fields remaining massless in its background; new maximal branch operators therefore enter the spectrum upon gauging. 
Anomaly matching is consistent with this picture: the branch sigma model carries no 1-form anomaly of its own, so the entire self-anomaly is accounted for by the frozen TQFT sectors.
A brane-move argument finally shows that at most two maximal branches of a given theory can be affected at the same time, and within the class of brane systems analysed here all three possibilities --- no, one, or two affected maximal branches --- are realised.
To encode the affected branches in the CSM magnetic quiver programme initiated in~\cite{Marino:2025uub}, an extension by non-simply laced edges has been proposed, with one edge of multiplicity $\kappa_i$ for each stack of $(1,\kappa_i)$ 5-branes. 
Within the analysed cases, the extended magnetic quiver makes the $\Z_g^{(1)}$ 1-form symmetry manifest; a mixed anomaly with the maximal branch isometry may be produced and, when present, is matched to that of the CSM theory; moreover, after gauging $\Gamma^{(1)}$, the magnetic quiver for the new theory is obtained.

\paragraph{Outlook.}
Beyond the class of Chern--Simons-Matter theories considered here, there exist several other classes of CSM theories whose symmetry structures deserve future analysis. 
A coarse classification separates these into: (i) CSM quiver theories that originate as world-volume theories of some Type II brane configuration, and (ii) more exotic theories that have no known brane realisation.

Naturally, brane configurations appear to be the more accessible next step. Specifically for orthosymplectic CSM quivers, some results are already available: for instance, in \cite{Bhardwaj:2022maz,Bergman:2024its,Harding:2025vov} symmetries and anomalies have been discussed; while, \eg \cite{Marino:2025ihk} offers a unified perspective on the maximal branches. 
Moreover, the techniques developed in this paper are expected to probe circular unitary CSM theories as well, where, however, the presence of additional $\urm(1)$ global symmetry factors should lead to extra features in the symmetry structure.

\paragraph{Acknowledgements.}
The authors thank Matthew Buican, Andrea Ferrari, Francesco Mignosa, Sinan Moura Soysüren, and Leonardo Santilli for interesting discussions.
The work of FM, FP and MS is supported by the Austrian Science Fund (FWF), START project ``Phases of quantum field theories: symmetries and vacua'' STA 73-N [grant DOI: 10.55776/STA73]. FM, FP and MS also acknowledge support from the Faculty of Physics, University of Vienna. FP acknowledges the financial support by the Vienna Doctoral School in Physics (VDSP).

\appendix
\section{\texorpdfstring{Dressed monopole operators in linear $\Ncal=4$ CSM quivers}{}}
\label{app:dressed_monopoles}
Dressed monopole operators in Chern--Simons matter theories have been studied extensively, \eg\ \cite{Bashkirov:2010kz,Aharony:2015pla,Cremonesi:2015dja,Cremonesi:2016nbo,Assel:2018wtj,Bergman:2020ifi}. The representation-theoretic ingredients needed to determine when a bare monopole operator in the quivers of the main text admits a gauge-invariant dressing, and when this dressing can be realised by fields that remain massless in the monopole background, are collected in this appendix.

\subsection{Dressing and maximal branch criteria}
\label{app:dressed_monopoles_preliminaries}
For a $\urm(N_s)$ gauge node, a magnetic flux is a coweight
\begin{align}
    m^{(s)}=(m^{(s)}_1,\ldots,m^{(s)}_{N_s})\in\Z^{N_s}\,,
    \qquad
    m^{(s)}_1\geq\cdots\geq m^{(s)}_{N_s}\,.
    \label{eq:dominant_coweight}
\end{align}
An irreducible gauge representation whose highest weight is proportional to $m^{(s)}$ is assigned to the corresponding bare monopole by the Chern--Simons coupling. If all components of a dominant weight $\mu$ are non-negative, the associated representation is $\mathcal R^{\urm(N)}_\mu\cong\mathcal S_\mu(\mathbf N)$. If they are non-positive, it is equivalently written as $\mathcal R^{\urm(N)}_\mu\cong\mathcal S_{-\widetilde\mu}(\mathbf N^*)$, where $-\widetilde\mu=(-\mu_N,\ldots,-\mu_1)$. For mixed-sign weights, the following decomposition is introduced:
\begin{align}
    m^{(s,+)}_i\coloneqq\max(m^{(s)}_i,0)\,,
    \qquad
    m^{(s,-)}_i\coloneqq\min(m^{(s)}_i,0)\,.
    \label{eq:non-neg_non-pos_part}
\end{align}
The representation $\mathcal R_{\kappa m^{(s)}}$ is then the highest-weight summand of $\mathcal R_{\kappa m^{(s,+)}}\otimes\mathcal R_{\kappa m^{(s,-)}}$. A sufficient criterion for mixed-sign dressings is thereby provided by treating the two sign sectors separately; it is not asserted to be necessary in full generality.

For a bifundamental hypermultiplet, the two chiral multiplets transform as $Q\in\mathbf N_1\otimes\mathbf N_2^*$ and $\widetilde Q\in\mathbf N_1^*\otimes\mathbf N_2$. Their symmetric products are decomposed by Cauchy's identity,
\begin{subequations}
\label{eq:Schur_decomp_hyper}
\begin{align}
 \mathrm{Sym}^A(\mathbf N_1\otimes\mathbf N_2^*)&\cong
 \bigoplus_\rho\mathcal S_\rho(\mathbf N_1)\otimes\mathcal S_\rho(\mathbf N_2^*)\,,\\
 \mathrm{Sym}^B(\mathbf N_1^*\otimes\mathbf N_2)&\cong
 \bigoplus_\mu\mathcal S_\mu(\mathbf N_1^*)\otimes\mathcal S_\mu(\mathbf N_2)\,.
\end{align}
\end{subequations}
Two consequences will be used throughout, cf.\ \cite{Fulton:2004uyc}:
\begin{enumerate}[label=(S\arabic*),ref=S\arabic*,leftmargin=2.5em]
    \item \label{item:orthogonality}\emph{Orthogonality.} The product $\mathcal S_\lambda(V)\otimes\mathcal S_\mu(V^*)$ contains a $\urm(\dim V)$ singlet iff $\lambda=\mu$, after zero-padding to a common length.
    \item \label{item:vanishing}\emph{Vanishing.} If $\ell_+(\mu)\coloneqq\#\{i:\mu_i>0\}$, then $\mathcal S_\mu(V)=0$ whenever $\ell_+(\mu)>\dim V$.
\end{enumerate}
Gauge invariance is therefore reduced to matching the partitions supplied by the bare monopole and by its dressing. A second condition is imposed by being a maximal branch operator. A bifundamental component of weight $e_i^{(s)}-e_j^{(s+1)}$ is massless precisely when
\begin{align}
    (e_i^{(s)}-e_j^{(s+1)})(m^{(s)},m^{(s+1)})
    =m_i^{(s)}-m_j^{(s+1)}=0\,.
    \label{eq:massless_bifundamental_criterion}
\end{align}
Thus, a gauge-invariant monopole is on a given maximal branch only if its required dressing can be built from flux-matched components along the corresponding links.

\subsection{\texorpdfstring{Two-node quiver: $\urm(N_1)_{\kappa}\times \urm(N_2)_{-\kappa}$}{Two-node quiver}}
\label{app:monopole_CSM_2-node}
The two-node theory with $\kappa>0$ is considered first,
\begin{align}
\label{eq:def_2-node_ex}
 \raisebox{-.5\height}{
    \includegraphics[]{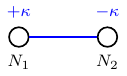}
    }
\end{align}
and dominant coweights $m^{(1)}$ and $m^{(2)}$. The bare monopole $\mathfrak M_{m^{(1)},m^{(2)}}$ is assigned to the representation
\begin{align}
    \mathcal R^{N_1}_{\kappa m^{(1)}}\otimes
    \left(\mathcal R^{N_2}_{\kappa m^{(2)}}\right)^*\,.
    \label{eq:rep_bare_mono}
\end{align}

\subsubsection{Integer coweights}
\label{app:monopole_CSM_2-node_integer}
The numbers of $Q_i{}^j$ and $\widetilde Q^{\,i}{}_j$ fields in a dressing are denoted by $a_{ij}$ and $b_{ij}$, respectively. The following conditions are required by gauge invariance:
\begin{align}
\begin{cases}
 \kappa m_i^{(1)}+\sum_j(a_{ij}-b_{ij})=0\,,\\
 -\kappa m_j^{(2)}-\sum_i(a_{ij}-b_{ij})=0\,,
\end{cases}
\qquad\Longrightarrow\qquad
\sum_i m_i^{(1)}=\sum_j m_j^{(2)}\,.
\label{eq:dress_physics}
\end{align}
For sign-definite coweights, this total-charge condition is strengthened by Schur orthogonality to componentwise matching, with zero-padding fixed by dominance. Assuming $N_1\geq N_2$,
\begin{subequations}
\begin{align}
 m^{(1)}&=(m^{(2)},0,\ldots,0) &&\text{for non-negative coweights}\,,
 \label{eq:matching_flux_positive}\\
 m^{(1)}&=(0,\ldots,0,m^{(2)}) &&\text{for non-positive coweights}\,.
 \label{eq:matching_flux_negative}
\end{align}
\end{subequations}
For mixed-sign coweights, the decomposition into $m^{(s,+)}$ and $m^{(s,-)}$ gives the sufficient condition
\begin{align}
 m^{(1)}=(m^{(2,+)},0,\ldots,0)+(0,\ldots,0,m^{(2,-)})\,.
 \label{eq:matching_flux_mixed}
\end{align}
The case $N_2>N_1$ follows by exchanging the nodes.

Equal non-zero flux components are paired by the matter fields used in these dressings. Indeed,
\begin{align}
    \rho_i{}^j(m^{(1)},m^{(2)})=m_i^{(1)}-m_j^{(2)}\,.
    \label{eq:mass_matter}
\end{align}
For non-negative coweights the required $\widetilde Q$ components lie on the matched indices, whereas for non-positive coweights the required $Q$ components lie on the correspondingly shifted matched indices. The two constructions may be combined for the mixed-sign sectors captured by \eqref{eq:matching_flux_mixed}. Hence the gauge-invariant sectors obtained above admit representatives dressed entirely by massless bifundamental fields and contribute to the maximal A branch.

\subsubsection{Fractional coweights}
\label{app:frac_flux_2-node}
The maximal anomaly-free subgroup of the $\Z_\kappa^{(1)}$ 1-form symmetry is $\Z_h^{(1)}$, where
\begin{align}
    h\coloneqq\gcd(N_1-N_2,\kappa)\,.
    \label{eq:def_g_2-node_CSM}
\end{align}
After this subgroup is gauged, the magnetic sectors have a common fractional shift
\begin{align}
    m_i^{(1)}=\widetilde m_i^{(1)}+\frac ah\,,
    \qquad
    m_j^{(2)}=\widetilde m_j^{(2)}+\frac ah\,,
    \qquad a\in\{0,\ldots,h-1\}\,.
\end{align}
The total charge constraint becomes
\begin{align}
    \sum_i\widetilde m_i^{(1)}-\sum_j\widetilde m_j^{(2)}
    =\frac ah(N_2-N_1)\,.
    \label{eq:charge_cancellation_fractional_final}
\end{align}
For $N_1=N_2$, the component-wise solution
\begin{align}
    m_i^{(1)}=m_i^{(2)}\,,
    \label{eq:frac_flux_2-node_CSM}
\end{align}
is allowed by Schur orthogonality, and the required dressing remains massless. Additional maximal A branch operators are therefore supplied by the fractional monopoles.

For unequal ranks, a non-trivial fractional shift cannot be accommodated by the zero-padding used in the sign-definite sectors. Gauge-invariant solutions may instead occur in mixed-sign sectors. For $N_1>N_2$, a representative family is
\begin{align}
 \widetilde m^{(1)}&=(0,\ldots,-(a-\ell)|\Delta N|)\,,
 &\widetilde m^{(2)}&=(\ell|\Delta N|,0,\ldots,0)\,,\\
 \Delta N&=\frac{N_2-N_1}{h}<0\,,
 \label{eq:frac_flux_2-node_N1>N2}
\end{align}
with $\ell\in\{0,\ldots,a\}$. The bifundamental masses are
\begin{align}
 \rho_i{}^j=-(a-\ell)|\Delta N|\delta_{i,N_1}
              -\ell|\Delta N|\delta_{j,1}\,.
\end{align}
The non-zero CS charge is supported precisely on the exceptional row or column for which no required bifundamental component is massless. These additional fractional sectors therefore do not furnish maximal A branch operators. The case $N_2>N_1$ is analogous.

\subsection{\texorpdfstring{Three-node prototype}{Three-node prototype}}
\label{app:monopole_CSM_3-node-1}
The three-node quiver
\begin{align}
\label{eq:def_3-node_ex1}
 \raisebox{-.5\height}{
    \includegraphics{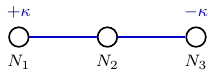}
    }
\end{align}
with Chern--Simons levels $(\kappa,0,-\kappa)$ is considered first. The bare monopole is assigned to the representation
\begin{align}
 \mathcal R^{\urm(N_1)}_{\kappa m^{(1)}}\otimes\mathbf 1^{\urm(N_2)}
 \otimes\left(\mathcal R^{\urm(N_3)}_{\kappa m^{(3)}}\right)^*\,.
 \label{eq:rep_bare_mono_3nodes_1}
\end{align}

\subsubsection{Integer coweights}
\label{app:monopole_CSM_3-node_ex1_integer}
Fluxes of the form
\begin{align}
 (m^{(1)},m^{(2)},m^{(3)})=(0,m^{(2)},0)
 \label{eq:B-branch_flux_3-nodes_ex1}
\end{align}
are gauge invariant without bifundamental dressing and give B branch operators, together with their dressings by Casimir invariants \cite{Cremonesi:2013lqa}.

For monopoles charged at the end nodes, the Chern--Simons charges must be cancelled jointly by the two bifundamental links. In the non-negative sector, the following conditions are first required by Schur vanishing:
\begin{align}
 N_2\geq\ell_+(m^{(1)})\,,
 \qquad
 N_2\geq\ell_+(m^{(3)})\,.
 \label{eq:constraint_N2_3-node-ex1}
\end{align}
The coweights $m^{(1)}$ and $m^{(3)}$ are then matched componentwise by Schur orthogonality, up to zero-padding when $N_1\neq N_3$,
\begin{subequations}
\label{eq:sol_non-neg_3-node_ex1}
\begin{align}
 N_1=N_3:&\quad m^{(1)}=m^{(3)}\,,\\
 N_1>N_3:&\quad m^{(1)}=(m^{(3)},0,\ldots,0)\,.
\end{align}
\end{subequations}
The non-positive sector is obtained by padding in front, and the mixed-sign criterion of the previous subsection may be applied to the two sign sectors separately. The central coweight $m^{(2)}$ is not fixed by gauge invariance.

Maximal A branch dressings must be constructed from components that remain massless on both links. Their fluxes are constrained by
\begin{align}
    m_i^{(1)}=m_j^{(2)}=m_k^{(3)}\,.
    \label{eq:flux_matching_chain}
\end{align}
Whenever \eqref{eq:constraint_N2_3-node-ex1} is satisfied, the central flux can be chosen so that every end-node component required by the dressing is transmitted through such a massless chain. A flux-independent sufficient condition is therefore
\begin{align}
    N_2\geq\min(N_1,N_3)\,.
    \label{eq:uniform_cond_3-node_ex1}
\end{align}

\subsubsection{Fractional coweights}
\label{app:gauging_1form_3-node-Ex1}
The anomaly-free 1-form symmetry is $\Z_g^{(1)}$, with $g=\gcd(N_1-N_3,\kappa)$. Charges are induced at both end nodes by a fractional shift of the B branch sectors \eqref{eq:B-branch_flux_3-nodes_ex1}, and dressing on both links is therefore required. For a non-trivial central integer flux, the two required link components are massive. No additional B branch operator is consequently produced.

\subsubsection{Fractional coweights: maximal branch}
\label{app:gauging_1form_maxBranch_3-node-Ex1}
For the A branch, fractionally shifted end-node coweights can be matched only when $N_1=N_3$, while the intermediate node must satisfy $N_2\geq N_1$. Once the good-node condition $2N_2\leq N_1+N_3$ \cite{Gaiotto:2008ak} is also imposed, only
\begin{align}
    N_1=N_2=N_3
\end{align}
is left as the case in which additional fractional monopoles furnish maximal A branch operators. Gauge-invariant mixed-sign solutions analogous to \eqref{eq:frac_flux_2-node_N1>N2} may still be found for unequal end ranks, but their dressings cannot be realised by massless fields.

\subsubsection{Generalisation}
\label{app:3-node_CSM_ex1_generalised}
The same argument may be applied to
\begin{align}
 \urm(N_1)_\kappa\times
 \prod_{r=1}^{n}\urm(N_{r+1})\times
 \urm(N_{n+2})_{-\kappa}\,.
 \label{eq:3-node_ex1_generalised}
\end{align}
The integer B branch fluxes are set to zero at the two end nodes and are arbitrary at the Chern--Simons-neutral nodes. No bifundamental dressing is required. After a common fractional shift, either the end-node charge constraint is not satisfied or massive dressing is required; hence no additional B branch operator is obtained.

For the A branch, the following condition is required by Schur vanishing:
\begin{align}
 \ell_+(m^{(1)}),\ \ell_+(m^{(n+2)})
 \leq\min_{1\leq r\leq n}N_{r+1}\,.
 \label{eq:size_constraint_n+2-node}
\end{align}
Whenever the corresponding gauge-invariant sector exists, the intermediate fluxes can be chosen to propagate every required end-node component through a massless chain. The flux-independent sufficient condition is
\begin{align}
 N_{r+1}\geq\min(N_1,N_{n+2})
 \qquad\text{for every }r\in\{1,\ldots,n\}\,.
\end{align}

For fractional sectors, equal end ranks and intermediate ranks no smaller than them are first required by maximal A branch dressing. Once the good-node inequalities $2N_{r+1}\leq N_r+N_{r+2}$ are imposed successively from the two endpoints towards the centre, all ranks are forced to be equal. Consequently,
\begin{align}
    N_1=N_2=\cdots=N_{n+2}
\end{align}
is required for the fractional gauge-invariant monopoles identified above to furnish maximal A branch operators. If the end ranks are unequal, additional mixed-sign gauge-invariant sectors may exist, but their dressing is not massless.

\subsection{Remark on alternating levels}
\label{app:monopole_CSM_3-node-2}

For the alternating-level quiver
\begin{align}
    \urm(N_1)_{\kappa}
    \times
    \urm(N_2)_{-\kappa}
    \times
    \urm(N_3)_{\kappa}\,,
\end{align}
the two bifundamental links are of opposite $\Ncal=4$ type. A maximal
A branch monopole must therefore be dressed using the first link only.
For integer coweights, this requires $m^{(3)}=0$, while the remaining
flux-matching condition is precisely the two-node condition of
Section~\ref{app:monopole_CSM_2-node}. Similarly, a maximal B branch
monopole requires $m^{(1)}=0$ and reduces to the two-node condition on
the second link.

The anomaly-free 1-form symmetry is $\Z_h^{(1)}$, with
\begin{align}
    h=\gcd(N_1-N_2+N_3,\kappa)\,.
\end{align}
After it is gauged, a non-trivial fractional sector is represented by
\begin{align}
    m_i^{(s)}
    =
    \widetilde m_i^{(s)}+\frac{a}{h}\,,
    \qquad
    a\in\{1,\ldots,h-1\}\,.
\end{align}
A vanishing coweight at either end node is then impossible. Non-trivial
CS gauge charges are consequently induced at both end
nodes, and matter fields from both links are required for their
cancellation. Since the two links are of opposite $\Ncal=4$ type,
such a dressing cannot furnish an operator on either maximal branch.
Thus, no additional maximal A or B branch monopole operator is
generated by the fractional sectors. Gauge-invariant dressed
monopoles away from these maximal branches are not excluded by this
argument.

\subsection{\texorpdfstring{Dressed monopoles in $\Ncal=3$ CSM quivers}{Dressed monopoles in N=3 CSM quivers}}
\label{app:dressed_monopoles_N3}
The two-node analysis may also be applied to the $\Ncal=3$ quiver
\begin{align}
    \urm(N_1)_{\kappa_1}\times\urm(N_2)_{-\kappa_2}\,,
    \qquad \kappa_1,\kappa_2>0\,.
\end{align}
For non-negative coweights, agreement of the weighted fluxes is required by Schur orthogonality. Thus,
\begin{subequations}
\begin{align}
 N_1=N_2:&\quad \kappa_1m_i^{(1)}=\kappa_2m_i^{(2)}\,,\\
 N_1>N_2:&\quad
 \kappa_1m_i^{(1)}=\kappa_2m_i^{(2)}\ \ (i\leq N_2)\,,
 \qquad m_i^{(1)}=0\ \ (i>N_2)\,.
\end{align}
\end{subequations}
The corresponding statements for non-positive and mixed-sign coweights follow by the same padding prescriptions as above. On a matched component, however,
\begin{align}
 \rho_i{}^i(m^{(1)},m^{(2)})
 =m_i^{(1)}-m_i^{(2)}
 =m_i^{(1)}\left(1-\frac{\kappa_1}{\kappa_2}\right)\,.
\end{align}
The dressing fields are therefore generically massive when $\kappa_1\neq\kappa_2$, and the corresponding gauge-invariant dressed monopoles cannot be identified as maximal branch operators. Since this obstruction is already present for integer coweights, the required masslessness is not restored by a common fractional shift, and the fractional case is not treated separately.
\section{\texorpdfstring{Details on CSM version of $T[\surm(N)]$}{Details on CSM version of TSUN}}
\label{app:CSM_TSUN}
Consider the brane system \eqref{eq:example_4_Unitary} with the CSM quiver
\begin{equation}
    \raisebox{-.5\height}{\includegraphics[width=\linewidth]{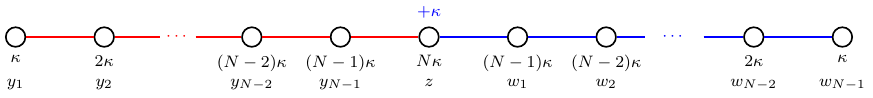}}
    \label{eq:CSM_TSUN}
\end{equation}
where below the gauge ranks the assigned topological fugacities $\{y_i\}_{i=1}^{N-1}$, $z$, $\{w_j\}_{j=1}^{N-1}$ are shown. The magnetic fluxes are denoted by $\{m^{(i)}_j\}_{j=1}^{n_i}$ for the $i$-th node of rank $n_i$.

\paragraph{0-form symmetry.}
By standard arguments, one finds the $\psurm(N)_A\times\psurm(N)_B$ currents. For concreteness, begin with the A branch. The simple roots $\alpha_i$ of $\surmL(N)_A$ are realised by monopole operators with non-trivial fluxes only at the $i$-th gauge node (counting from the left) for $i\in\{1,2,\ldots,N-1\}$; \ie $m^{(i)}=(1,0,\ldots,0)$ and $m^{(j)}=(0,\ldots,0)$ for all $j\neq i$. 
For $\surmL(N)_B$, the simple roots are realised via monopoles with non-trivial fluxes $m^{(N+j)}=(1,0,\ldots,0)$ only at the $N+j$-th gauge node for $j\in \{1,2,\ldots, N-1\}$.

In terms of fugacity maps, one employs the standard Cartan matrix map between root space fugacities $y_i$ (resp.\ $w_i$) and weight space fugacities $a_i$ (resp.\ $b_i$)
\begin{align}
    y_i \mapsto \prod_{j=1}^N a_j^{C_{ij}^{A_{N-1}}}\;, \qquad 
        w_i \mapsto \prod_{j=1}^N b_j^{C_{ij}^{A_{N-1}}} \;,
        \label{eq:fugacity_map_CSM_TSUN}
\end{align}
where $C_{ij}^{A_{N-1}}$ is the $A_{N-1}$ Cartan matrix. With this, the index expansion yields 
\begin{align}
    I = 1 + \left( \chi_{[1,0,\ldots,0,1]}(b_i)\, t^{2} + \chi_{[1,0,\ldots,0,1]}(a_i)\, t^{-2}  \right) x + \ldots \,,
\end{align}
wherein $t$ denotes the axial fugacity.

\paragraph{Wilson lines and 1-form symmetry.}
Next, consider the $\urm(N\kappa)$ fundamental Wilson line $W$. Since it cannot end, one also finds that $W^{\otimes r}$ insertions in the index yield vanishing expectation values for all $r< \kappa$.  Most interesting are the monopole operators on which $W^{\otimes \kappa}$ can end. One finds that the flux at the central $\urm(N\kappa)$ node has to be set to $m^{(N)}=(0,\ldots,0,-1)\equiv (\circ,-1)$, simply to cancel the gauge charges of $W^{\otimes \kappa}$. However, non-trivial fluxes at the other gauge nodes can be turned on as well and contribute at the same order, provided the following configurations are chosen:
\begin{align}
\mathfrak{M}_{(r,s)} \equiv
\begin{cases}
     m^{(i)} = (\circ,-1) &\text{for } r\leq i < N  \\
    m^{(N)} = (\circ,-1) &  \\    
    m^{(j)} = (\circ,-1) &\text{for } N< j \leq s  
\end{cases}
\end{align}
for any $r\in \{1,\ldots, N\}$ and any $s\in\{N,\ldots,2N-1\}$. Overall, this gives rise to $N^2$ monopole operators $\mathfrak{M}_{(r,s)}$ on which $W^{\otimes \kappa}$ can end. This allows one to compute the representation content of the flavour Wilson line that $W^{\otimes \kappa}$ is equivalent to. Using the topological charges of the $\mathfrak{M}_{(r,s)}$ together with the fugacity map \eqref{eq:fugacity_map_CSM_TSUN}, one finds 
\begin{align}
    \mathfrak{M}_{(r,s)} \sim \prod_{i=r}^{N-1} y_i\cdot z \cdot \prod_{j=1}^{s-N} w_j
    = 
    \begin{cases} 
        \frac{a_r}{a_{r-1}} a_{N-1}\ z\ b_1 \frac{b_s}{b_{s-1}} & r>1, s<N-1  \,,\\
         a_1 a_{N-1}\ z\ b_1 \frac{b_s}{b_{s-1}} & r=1, s<N-1\,,\\
        \frac{a_r}{a_{r-1}} a_{N-1}\ z\ b_1 b_{N-1} & r>1, s=N-1 \,,\\
       a_1 a_{N-1}\ z\ b_1 b_{N-1} & r=1, s=N-1 \,.
    \end{cases}
\end{align}
Inspecting this motivates a final fugacity map 
\begin{align}
    z \mapsto \frac{Q}{a_{N-1} b_1}
\end{align}
such that 
\begin{align}
    \mathfrak{M}_{(r,s)} \sim
   Q \begin{cases} 
        \frac{a_r}{a_{r-1}}\ \frac{b_s}{b_{s-1}} & r>1, s<N-1  \,,\\
         a_1 \ \frac{b_s}{b_{s-1}} & r=1, s<N-1\,,\\
        \frac{a_r}{a_{r-1}} \ b_{N-1} & r>1, s=N-1 \,,\\
       a_1 \ b_{N-1} & r=1, s=N-1 \,.
    \end{cases}
\end{align}
and all these monopoles together furnish the following representation
\begin{align}
    \sum_{r,s}\mathfrak{M}_{(r,s)} \sim Q  \cdot \chi_{[1,0,\ldots,0]}(a_i)\cdot \chi_{[0,\ldots,0,1]}(b_i)  \,.
\end{align}
In other words, the unit anti-monopole $(\circ,-1)$ at the CS-node $\urm(N\kappa)$ transforms as $[1,0,\ldots,0]_A \otimes [0,\ldots,0,1]_B$. Consequently, the unit-monopole $(1,\circ)$ at the CS-node $\urm(N\kappa)$ transforms as $[0,\ldots,0,1]_A \otimes [1,0,\ldots,0]_B$.

\bibliographystyle{JHEP}  
\bibliography{references}

\endgroup

\end{document}